\documentclass[10pt,journal,compsoc]{IEEEtran}

\ifCLASSOPTIONcompsoc
  \usepackage[nocompress]{cite}
\else
  \usepackage{cite}
\fi

\usepackage{booktabs}
\usepackage[table,xcdraw]{xcolor}
\usepackage{amssymb}
\usepackage{graphicx}
\usepackage{multirow}
\usepackage{bbding}
\usepackage{subfig}

\ifCLASSINFOpdf
\else
\fi

\begin{document}

\title{CMKD: CNN/Transformer-Based Cross-Model Knowledge Distillation for Audio Classification}

\author{Yuan Gong,~\IEEEmembership{Member,~IEEE,}
        Sameer Khurana,
        Andrew Rouditchenko,
        and James Glass,~\IEEEmembership{Fellow, IEEE}
\thanks{Preprint. Code will be released after the review process.}
\thanks{The authors are with the Computer Science and Artificial Intelligence Laboratory, Massachusetts Institute of Technology, Cambridge, MA 02139 USA (e-mail: \{yuangong, skhurana, roudi, glass\}@mit.edu).}}

\markboth{Preprint}%
{Shell \MakeLowercase{\textit{et al.}}: Bare Demo of IEEEtran.cls for Computer Society Journals}

\IEEEtitleabstractindextext{%
\begin{abstract}
Audio classification is an active research area with a wide range of applications. Over the past decade, convolutional neural networks (CNNs) have been the de-facto standard building block for end-to-end audio classification models. Recently, neural networks based solely on self-attention mechanisms such as the Audio Spectrogram Transformer (AST) have been shown to outperform CNNs. 
In this paper, we find an intriguing interaction between the two very different models - \emph{CNN and AST models are good teachers for each other}. When we use either of them as the teacher and train the other model as the student via knowledge distillation (KD), the performance of the student model noticeably improves, and in many cases, is better than the teacher model. In our experiments with this CNN/Transformer \underline{C}ross-\underline{M}odel \underline{K}nowledge \underline{D}istillation (CMKD) method we achieve new state-of-the-art performance on FSD50K, AudioSet, and ESC-50.
\end{abstract}

\begin{IEEEkeywords}
Audio Classification, Convolutional Neural Networks, Transformer, Knowledge Distillation
\end{IEEEkeywords}}

\maketitle

\IEEEdisplaynontitleabstractindextext

\IEEEpeerreviewmaketitle
\IEEEraisesectionheading{\section{Introduction}
\label{sec:introduction}}

\IEEEPARstart{A}{udio} classification aims to identify sound events that occur in a given audio recording and enables a variety of artificial intelligence-based systems to disambiguate sounds and understand the acoustic environment. Historically, hand-crafted features and hidden Markov models (HMMs) were used for audio classification~\cite{woodard1992modeling,goldhor1993recognition,chachada2014environmental}.
With the rise of neural networks in the past decade, \textbf{Convolutional Neural Networks~(CNNs)}~\cite{lecun1995convolutional} have became the de-facto standard building block for \emph{end-to-end} audio classification models, which aim to learn a direct mapping from audio waveforms or spectrograms to corresponding labels~\cite{hershey2017cnn,palanisamy2020rethinking,salamon2017deep,piczak2015environmental,tokozume2017learning,huzaifah2017comparison}. More recently, neural networks based purely on self-attention, such as the \textbf{Audio Spectrogram Transformer (AST)}~\cite{gong21b_interspeech,koutini2021efficient,chen2022hts}, have been shown to further outperform deep learning models constructed with convolutional neural networks on various audio classification tasks, thus extending the success of Transformers from natural language processing~\cite{vaswani2017attention,devlin-etal-2019-bert} and computer vision~\cite{dosovitskiy2020image,touvron2021training} to the audio domain. 

CNN and Transformer models both have their advantages. For example, CNN models have several built-in inductive biases such as spatial locality and translation equivariance, making them well suited to spectrogram-based end-to-end audio classification. Transformer models do not have such built-in inductive biases and learn in a more data-driven manner, making them more flexible. As a consequence, the representations learned by CNN and Transformer models are noticeably different~\cite{raghu2021vision}. On the other hand, while Transformer models perform better, they are less computationally efficient than CNN models on long audio input due to their $O(n^2)$ complexity.

In this paper, we show an intriguing interaction between the two very different models - \emph{CNN and AST models are good teachers for each other}. When we use one model as the teacher and train another model as the student via knowledge distillation (KD), the performance of the student model noticeably improves, and in most cases, is better than the teacher model. We refer to this knowledge distillation framework between a CNN and a Transformer model as \textbf{\underline{C}ross-\underline{M}odel \underline{K}nowledge \underline{D}istillation (CMKD)}. The success of cross-model knowledge distillation is not trivial because 1) cross-model knowledge distillation works bi-directionally in both  \texttt{CNN}$\rightarrow$\texttt{Transformer} and \texttt{Transformer}$\rightarrow$\texttt{CNN} directions. Usually in KD, the teacher needs to be stronger than the student, but for cross-model knowledge distillation, a weak teacher can still improve a student's performance. 2) For both directions, the student outperforms the teacher after knowledge distillation, even when the teacher is originally stronger. 3) KD between two models of the same class leads to a much smaller or no performance improvement. Consequently, with the proposed cross-model knowledge distillation, a simple EfficientNet KD-CNN model with mean pooling outperforms the much larger AST model on FSD50K and ESC50 dataset. Conversely, the KD-AST model achieves new state-of-the-art performance on FSD50K, AudioSet, and ESC50 datasets. 

The contribution of this paper is threefold:
First, to the best of our knowledge, we are the first to explore bi-directional knowledge distillation between CNN and Transformer models; previous efforts~\cite{touvron2021training,ren2021co} only study the \texttt{CNN}$\rightarrow$\texttt{Transformer} direction and are performed in the visual domain. Second, we conduct extensive experiments on standard audio classification datasets and find the optimal knowledge distillation setting. In addition, we conduct a series of probing tests and ablation studies to explore the working mechanism of cross-model knowledge distillation. Third, due to the proposed cross-model KD framework, the small and efficient CNN models match or outperform previous state-of-the-art models; the AST models achieve even better performance and achieve new state-of-the-art results on FSD50K, AudioSet, and ESC50. We also evaluate CMKD on various CNN, CNN-attention, and Transformer models and find it improves the performance of all these models, showing CMKD is a generic method - all types of models can benefit from CMKD.

\begin{figure}[t]
  \centering
  \includegraphics[width=8.7cm]{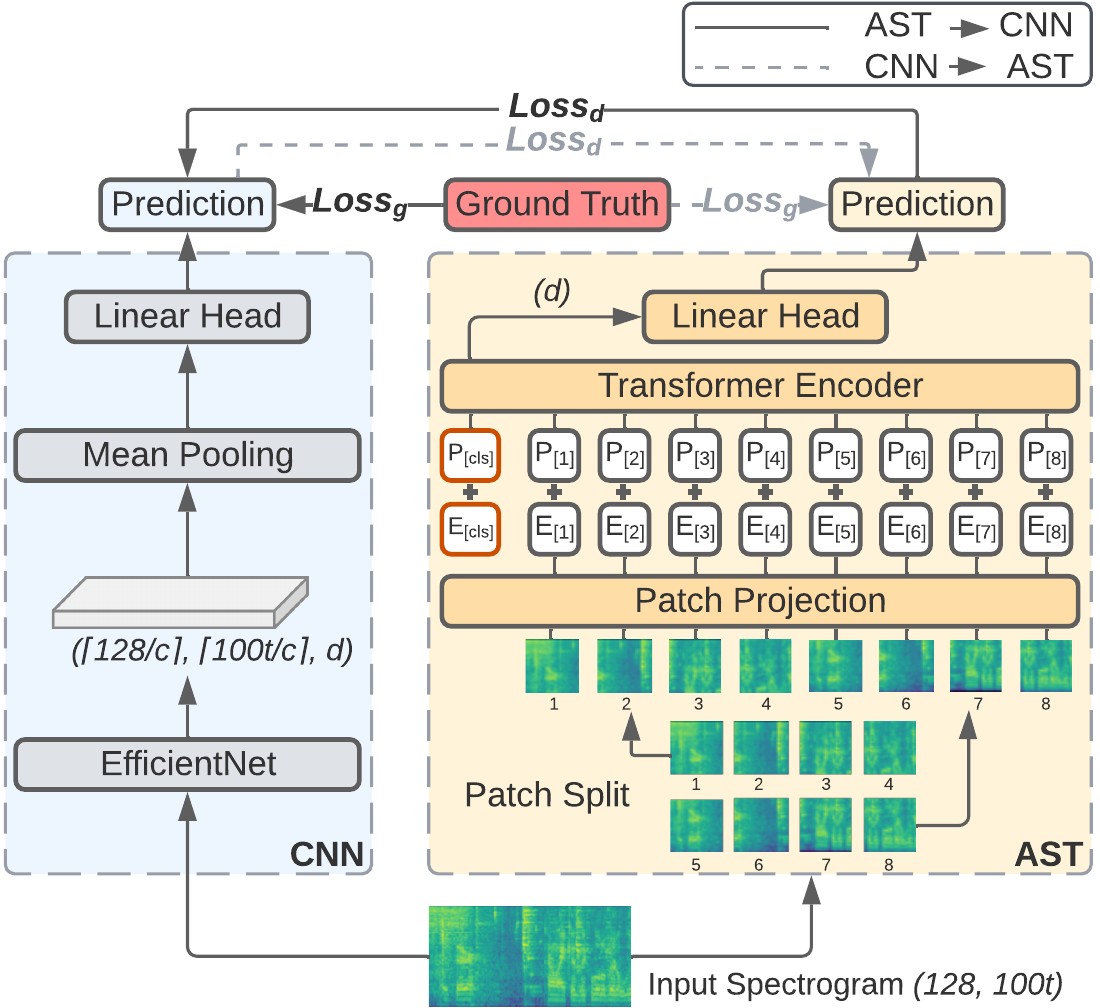}
  \caption{Illustration of the CNN and AST model architecture and the cross-model knowledge disitllation process. For both \texttt{AST}$\rightarrow$\texttt{CNN} and \texttt{CNN}$\rightarrow$\texttt{AST}, we always use a trained teacher model and freeze it during the student model training. The training objective is minimizing the ground truth loss $Loss_g$ and the distillation loss $Loss_d$.}
  \label{fig:ilus}
\end{figure}

\section{Cross-Model Knowledge Distillation}
\label{sec:method}

This section discusses the details of the proposed cross-model knowledge distillation between a CNN model and an AST model. Specifically, we introduce the architecture of CNN and AST models used in this work in Sections~\ref{sec:cnn_arc} and~\ref{sec:ast_arc}, respectively. We intend to use CNN and AST models that achieve the best performance on audio classification~\cite{gong_psla,gong21b_interspeech} to see if CMKD can further improve their performance and achieve a new state-of-the-art. We then discuss the main difference between these two classes of models in Section~\ref{sec:diff}. Finally, we show the knowledge distillation setting and notation in Section~\ref{sec:kd}.

\subsection{Convolution Neural Networks}
\label{sec:cnn_arc}

In this work, we use the CNN model without attention module proposed in~\cite{gong_psla}. To the best of our knowledge, it is the best CNN model on the audio classification task. As illustrated in the left side of Figure~\ref{fig:ilus}, the input audio waveform of~$t$ seconds is first converted into a sequence of~128-dimensional log Mel filterbank~(fbank) features computed with a~25ms Hanning window every~10ms. This results in a~$128\times100t$ spectrogram as input to the CNN model. The spectrogram is then input to a standard CNN model such as EffcientNet~\cite{tan2019efficientnet} and DenseNet~\cite{8099726}. The output of the penultimate layer is a tensor of size $(\lceil128/c\rceil, \lceil100t/c\rceil, d)$ in frequency, time, and embedding dimension, respectively, where $c$ is the feature downsampling factor of the CNN model. We apply a time and frequency mean pooling to produce the $d$ dimensional spectrogram-level representation. Finally, a linear layer with a sigmoid (for multi-label classification) or softmax (for single-label classification) activation maps the audio spectrogram representation to labels for classification.
We mainly use EfficientNet-B2~\cite{tan2019efficientnet} as the CNN model in our experiments because it has been shown to be the best CNN model on audio classification~\cite{gong_psla}. To study the impact of model size and the generalization of the proposed cross-model knowledge distillation strategy, we also include Efficient-B0, Efficient-B6, and DenseNet-121 in our experiments. The statistics of the different models are presented in Table~\ref{tab:models}.

\begin{table}[t]
\setlength\tabcolsep{3.5pt}
\centering
\caption{Statistics of the CNN and AST models used in this work.}
\begin{tabular}{@{}lcccc@{}}
\toprule
\multicolumn{1}{c}{Model} & \begin{tabular}[c]{@{}c@{}}\# Layers\\ /Blocks\end{tabular} & \begin{tabular}[c]{@{}c@{}}\# Attention\\ Heads\end{tabular} & \begin{tabular}[c]{@{}c@{}}Representation\\ Dimension $d$\end{tabular} & \begin{tabular}[c]{@{}c@{}}\# Params\\ (M)\end{tabular} \\ \midrule
\multicolumn{5}{l}{{\color[HTML]{656565} \textit{CNN Models}}}                                                                                                                                                                                \\
EfficientNet-B0           & 237                                                         & -                                                            & 1,280                                                                 & 4             \\
EfficientNet-B2           & 339                                                         & -                                                            & 1,408                                                                 & 8             \\
EfficientNet-B6           & 666                                                         & -                                                            & 2,304                                                                 & 41             \\
DenseNet-121              & 121                                                         & -                                                            & 1,024                                                                 & 7             \\ \midrule
\multicolumn{5}{l}{{\color[HTML]{656565} \textit{AST Models}}}                                                                                                                                                                        \\
AST-Tiny                  & 12                                                          & 3                                                            & 192                                                                  & 6             \\
AST-Small                 & 12                                                          & 6                                                            & 384                                                                  & 23             \\
AST-Base                  & 12                                                          & 12                                                           & 768                                                                  & 88             \\ \bottomrule
\end{tabular}
\label{tab:models}
\end{table}

\subsection{Audio Spectrogram Transformers}
\label{sec:ast_arc}

In this work, we use the original AST model proposed in~\cite{gong21b_interspeech} that has the best performance on the audio classification task. As illustrated on the right side of Figure~\ref{fig:ilus}, the input audio waveform of~$t$ seconds is converted to a~$128\times100t$ spectrogram in the same way as the CNN model.
We then split the spectrogram into a sequence of~$N$ $16\times16$ patches with an overlap of~6 in both time and frequency dimension, where $N=12\lceil{(100t-16)/10}\rceil$ is the number of patches and the effective input sequence length for the Transformer.
We flatten each~$16\times16$ patch to a 1D patch embedding of size $d$ using a linear projection layer.
We refer to this linear projection layer as the patch embedding layer.
Since the Transformer architecture does not capture input order information and the patch sequence is also not in temporal order, we add a trainable positional embedding~(also of size~$d$) to each patch embedding to allow the model to capture the spatial structure of the 2D audio spectrogram. We append a \texttt{[CLS]} token at the beginning of the sequence.
The resulting sequence is then input to a standard Transformer encoder~\cite{vaswani2017attention}. The Transformer encoder's output of the \texttt{[CLS]} token serves as the audio spectrogram representation. A linear layer with sigmoid activation (for multi-label classification) or softmax (for single-label classification) maps the audio spectrogram representation to labels for classification.  To study the impact of model size, we use three AST models of different sizes in our experiments: AST-Tiny, AST-Small, and AST-Base. The statistics of the models are shown in Table~\ref{tab:models}.

For both CNN and AST models (including No-KD baseline models), we always use ImageNet pretraining, a common technique to improve audio classification~\cite{gong_psla,gong21b_interspeech,gwardys2014deep,guzhov2020esresnet,adapa2019urban,palanisamy2020rethinking}. Specifically, we initialize the CNN and Transformer with ImageNet-pretrained weights from public model checkpoints \footnote{https://pytorch.org/vision/stable/models.html}$^,$\footnote{https://github.com/rwightman/pytorch-image-models} and then conduct audio classification training. Since the input to the vision models is a 3-channel image while the audio model input is a single-channel spectrogram, we average the weights corresponding to each of the three input channels of the vision model checkpoints and use them as the weights of the first layer of the audio models. This is equivalent to expanding a single-channel spectrogram to 3-channels with the same content, but is computationally more efficient. Note that for all AST models (including the No-KD baselines), we initialize the model with the same vision model checkpoint~\cite{touvron2021training} as the original AST~\cite{gong21b_interspeech}. This vision model checkpoint is trained with KD during the ImageNet training stage~\cite{touvron2021training} and has been shown to lead to better performance for audio classification than checkpoints trained without KD. This setting allows us to clearly identify the improvement due to KD in the audio domain training stage and make a fair comparison with the original AST~\cite{gong21b_interspeech}.

\subsection{Difference between CNN and AST}
\label{sec:diff}

As described in the previous two subsections, CNN and AST models have completely different design philosophies. CNN models have several built-in inductive biases that make them well suited to the spectrogram-based end-to-end audio classification task such as translation equivariance (without pooling) or translation invariance (with pooling) and locality. CNN models typically have a hierarchical architecture with an increasingly large receptive field in deeper layers. In contrast, AST lacks such spectrogram-specific built-in inductive biases and needs to learn in a more data-driven manner. However, this also allows AST models to be more flexible, e.g., the receptive field of an AST is learnable and the very first layer can capture long-range global context while the locality makes the first few layers of CNNs struggle to relate temporal- and frequency- distant features. As a consequence, CNNs typically perform better with small datasets while Transformer models start showing their superiority with more training data~\cite{liu2022convnet,trockman2022patches,touvron2021training,dosovitskiy2020image}. Due to the different model architectures, the learned representations of CNNs and Transformers have striking differences~\cite{raghu2021vision}. Efficiency-wise,  AST has quadratic complexity with respect to the input size due to the global attention design while the CNN model has linear complexity and thus is more efficient for long audio samples. 

\subsection{Knowledge Distillation}
\label{sec:kd}

In this work, we follow the original knowledge distillation setup~\cite{hinton2015distilling} with \emph{consistent teaching}~\cite{beyer2021knowledge} and find this simple setup works better than the more complex attention distillation strategy~\cite{touvron2021training} for the audio classification task. Specifically, we first train the teacher model, and then during the student model training, we feed the input audio spectrogram with the exact same augmentations (e.g., SpecAugment~\cite{park2019specaugment}, noise, time shift, and mixup~\cite{zhang2018mixup}) to the teacher and student models (consistent teaching). We use the following loss for the student model training:

\begin{equation}
    Loss = \lambda Loss_g(\psi(Z_s),y) + (1-\lambda) Loss_d(\psi(Z_s), \psi(Z_t/\tau))
\end{equation}

where $\lambda$ is the balancing coefficient; $Loss_g$ and $Loss_d$ are the ground truth and distillation losses, respectively; $\psi$ is the activation function; $Z_s$ and $Z_t$ are the logits of the student and teacher model, respectively; and $\tau$ is the temperature. The teacher model is frozen during the student model training.

By default, we use the Kullback–Leibler divergence as $Loss_d$. For cross-model KD, the teacher and student may have different logit distribution softness, thus we only apply $\tau$ on the teacher logits to explicitly control the difference. For simplicity, we fix $\lambda=0.5$ and do not scale it with $\tau$. We use the cross-entropy (CE) loss as $Loss_g$ and the softmax activation function for single-label classification tasks such as ESC-50; the binary cross-entropy (BCE) loss as $Loss_g$ and the sigmoid activation function for multi-label classification tasks such as FSD50K and AudioSet.  We study the performance impact of KD settings in Section~\ref{sec:kd_setting}. In this paper, we use \texttt{teacher}$\rightarrow$\texttt{student} to denote the direction of knowledge distillation, e.g., \texttt{EfficientNet-B0}$\rightarrow$\texttt{AST-Base} means we first train an EfficientNet-B0 model, and then use it as the teacher to train an AST-Base model as the student.


\begin{table}[]
\centering
\caption{Training settings of CNN and AST models on FSD50K, AudioSet, and ESC-50. Baseline and KD models are trained with same settings.}
\setlength\tabcolsep{3.5pt}
\begin{tabular}{@{}ccccccc@{}}
\toprule
Setting                                                                & \multicolumn{2}{c|}{FSD50K}      & \multicolumn{2}{c|}{AudioSet}    & \multicolumn{2}{c}{ESC-50}         \\ \midrule
                                                                       & CNN  & \multicolumn{1}{c|}{AST}  & CNN  & \multicolumn{1}{c|}{AST}  & CNN              & AST             \\ \cmidrule(l){2-7} 
Initial LR                                                             & 5e-4 & \multicolumn{1}{c|}{5e-5} & 1e-4 & \multicolumn{1}{c|}{1e-5} & 1e-4             & 1e-5            \\
Batch Size                                                             & 24   & \multicolumn{1}{c|}{12}   & 120  & \multicolumn{1}{c|}{12}   & \multicolumn{2}{c}{48}             \\
Epoch                                                                  & \multicolumn{2}{c|}{50}          & 30   & \multicolumn{1}{c|}{5}    & \multicolumn{2}{c}{25}             \\
Loss                                                                   & \multicolumn{4}{c|}{Binary Cross Entropy}                          & \multicolumn{2}{c}{Cross Entropy} \\
Mixup Ratio                                                            & \multicolumn{4}{c|}{0.5}                                            & \multicolumn{2}{c}{0}              \\
\begin{tabular}[c]{@{}c@{}}SpecAugment\\ (Frequency/Time)\end{tabular} & \multicolumn{4}{c|}{48/192}                                         & \multicolumn{2}{c}{24/96}          \\
Class Balancing                                                        & \multicolumn{4}{c|}{Yes}                                            & \multicolumn{2}{c}{No}             \\ \midrule\midrule
Label Smoothing                                                        & \multicolumn{6}{c}{0.1}                                                                                  \\
Random Noise                                                           & \multicolumn{6}{c}{$U(0, 0.05)$ on spectrogram}                                                                                  \\
Random Time Shift                                                      & \multicolumn{6}{c}{$\pm$10 frames}                                                                                  \\ \bottomrule
\end{tabular}
\label{tab:trsetting}
\end{table}

\begin{table*}[t]
\centering
\caption{\centering{MAP of the student model of various \texttt{(teacher,student)} pair on the FSD50K evaluation set. $\dagger$ denotes that student model performance improves with KD; $\ast$ denotes that student model outperforms teacher model.}}
\begin{tabular}{@{}ccclllllll@{}}
\toprule
\begin{tabular}[c]{@{}c@{}}Student\\ Model\end{tabular} & \begin{tabular}[c]{@{}c@{}}Model Size\\ (\# Params)\end{tabular} & \multicolumn{8}{c}{\begin{tabular}[c]{@{}c@{}}Teacher Model\\ (Teacher Model Performance Shown in Grey)\end{tabular}}                                                                                                                                                                                                                                                                             \\ \midrule
                                                        &                                                                  & \textit{No KD}                      & \multicolumn{1}{c}{EffNet-B0}                   & \multicolumn{1}{c}{EffNet-B2}                   & \multicolumn{1}{c}{EffNet-B6}                   & \multicolumn{1}{c}{DenseNet-121}                & \multicolumn{1}{c}{AST-Tiny}                    & \multicolumn{1}{c}{AST-Small}                   & \multicolumn{1}{c}{AST-Base}                    \\
                                                        &                                                                  & {\color[HTML]{656565} \textit{N/A}} & \multicolumn{1}{c}{{\color[HTML]{656565} 55.7}} & \multicolumn{1}{c}{{\color[HTML]{656565} 56.2}} & \multicolumn{1}{c}{{\color[HTML]{656565} 54.7}} & \multicolumn{1}{c}{{\color[HTML]{656565} 53.9}} & \multicolumn{1}{c}{{\color[HTML]{656565} 56.3}} & \multicolumn{1}{c}{{\color[HTML]{656565} 56.6}} & \multicolumn{1}{c}{{\color[HTML]{656565} 57.6}} \\ \cmidrule(l){3-10} 
EffNet-B0                                               & 4M                                                               & \textit{55.7}                       & \ \quad55.9$^{\dagger\ast}$                            & \ \quad56.0$^\dagger$                                 & \ \quad55.4$^\ast$                                    & \ \quad\quad56.1$^{\dagger\ast}$                       & \textbf{\ \quad57.4$^{\dagger\ast}$}        & \ \quad56.9$^{\dagger\ast}$                 & \ \quad56.6$^\dagger$                      \\
EffNet-B2                                               & 8M                                                               & \textit{56.2}                       & \ \quad56.2$^\ast$                                    & \ \quad56.3$^{\dagger\ast}$                            & \ \quad55.4$^\ast$                                    & \ \quad\quad56.8$^{\dagger\ast}$                       & \textbf{\ \quad58.2$^{\dagger\ast}$}        & \ \quad57.4$^{\dagger\ast}$                 & \ \quad57.3$^\dagger$                      \\
EffNet-B6                                               & 41M                                                              & \textit{54.7}                       & \ \quad55.6$^\dagger$                                 & \ \quad56.1$^\dagger$                                 & \ \quad54.0                                           & \ \quad\quad55.1$^{\dagger\ast}$                       & \textbf{\ \quad58.0$^{\dagger\ast}$}        & \ \quad57.7$^{\dagger\ast}$                 & \ \quad57.8$^{\dagger\ast}$                 \\
DenseNet-121                                            & 7M                                                               & \textit{53.9}                       & \ \quad53.1$  $                                       & \ \quad54.6$^\dagger$                                 & \ \quad54.0$^\dagger$                                 & \ \quad\quad52.7                                      & \textbf{\ \quad56.9$^{\dagger\ast}$}        & \ \quad56.7$^{\dagger\ast}$                 & \ \quad57.0$^\dagger$                      \\
AST-Tiny                                                & 6M                                                               & \textit{56.3}                       & \ \quad58.6$^{\dagger\ast}$                            & \ \quad58.5$^{\dagger\ast}$                            & \ \quad58.2$^{\dagger\ast}$                            & \textbf{\ \quad\quad59.1$^{\dagger\ast}$}              & \ \quad57.2$^{\dagger\ast}$                 & \ \quad56.5$^\dagger$                      & \ \quad56.8$^\dagger$                      \\
AST-Small                                               & 23M                                                              & \textit{56.6}                       & \textbf{\ \quad60.5$^{\dagger\ast}$}                   & \ \quad60.1$^{\dagger\ast}$                            & \ \quad59.3$^{\dagger\ast}$                            & \ \quad\quad60.3$^{\dagger\ast}$                       & \ \quad59.4$^\dagger$                      & \ \quad57.7$^{\dagger\ast}$                 & \ \quad57.8$^{\dagger\ast}$                 \\
AST-Base                                                & 88M                                                              & \textit{57.6}                       & \textbf{\ \quad61.7$^{\dagger\ast}$}                   & \ \quad61.5$^{\dagger\ast}$                            & \ \quad61.5$^{\dagger\ast}$                            & \ \quad\quad61.5$^{\dagger\ast}$                       & \ \quad60.9$^{\dagger\ast}$                 & \ \quad59.6$^{\dagger\ast}$                 & \ \quad59.6$^{\dagger\ast}$                 \\ \bottomrule
\end{tabular}
\label{tab:teacherstudent}
\end{table*}

\section{Experiments Settings}
\label{sec:exp_setting}

\subsection{Datasets}
\label{sec:dataset}

We evaluate the proposed method on three widely-used audio classification datasets: FSD50K~\cite{fonseca2020fsd50k}, AudioSet~\cite{gemmeke2017audio}, and ESC-50~\cite{piczak2015esc}. The FSD50K dataset~\cite{fonseca2020fsd50k} is a collection of sound event audio clips with 200 classes, which contains 37,134 audio clips for training, 4,170 audio clips for validation, and 10,231 audio clips for evaluation. The audio clips are of variable length from 0.3 to 30s with an average of 7.6s. In our experiments, we sample audio at 16kHz and trim all clips to 10s. We use the FSD50K dataset for the majority of our experiments in this paper (Section~\ref{sec:fsd50k}) for three reasons. First, FSD50K allows more rigorous experiments since it has an official training, validation, and evaluation split. In this paper, we train the model with the training set, tune all hyper-parameters and select the model based on the validation set, and report the mean average precision (mAP) on the evaluation set. Second, FSD50K is a publicly available dataset, so results on FSD50K are easy to reproduce. Third, the FSD50K dataset is of moderate size (50K samples) compared with AudioSet (2M samples) and ESC-50 (2K samples), which allows us to conduct extensive experiments with our computational resources. \footnote{In contrast, the AudioSet only has official training and evaluation split, and contains considerable label noise~\cite{shah2018closer,meire2019impact,fonseca2020addressing}. In addition, AudioSet is composed of YouTube videos that are not freely distributable. Downloading AudioSet from YouTube is laborious.}

AudioSet~\cite{gemmeke2017audio} is a collection of over 2 million 10-second audio clips excised from YouTube videos and labeled with the sounds that the clip contains from a set of 527 labels. The balanced training, full training, and evaluation set contain 22k, 2M, and 20k samples, respectively. We use AudioSet to study the generalization ability of the proposed method on larger datasets. The ESC-50~\cite{piczak2015esc} dataset consists of 2,000 5-second environmental audio recordings organized into 50 classes. We use ESC-50 to study the transferability of models trained with the proposed knowledge distillation method. 

\subsection{Training Settings}
\label{sec:tr_setting}

The hyper-parameters we use to train the models are shown in Table~\ref{tab:trsetting}. In order to make fair comparisons, we train models with and without KD with exactly the same settings. Specifically, we use the training recipes that are optimized for the original CNN and AST without KD in \cite{gong_psla,gong21b_interspeech} including the initial learning rate, learning rate scheduler, batch size, class balancing, mixup training~\cite{zhang2018mixup}, and SpecAugment~\cite{park2019specaugment}. This ensures the baseline models without KD are properly trained. The only changes we made from~\cite{gong_psla,gong21b_interspeech} is to add label smoothing, random noise augmentation, and random time shift augmentation to the training pipeline. These training techniques may have similar effects to cross-model knowledge distillation, e.g., label smoothing may have a similar effect to the teacher's soft label, random time shift augmentation may have a similar effect to transferring the inductive bias of translation invariance in \texttt{CNN}$\rightarrow$\texttt{AST}. 
We apply these training techniques to both baseline and KD models to exclude the improvement that can be achieved by such simple techniques, and thus more clearly identify the improvement from the cross-model knowledge distillation. In practice, we find these techniques slightly improve the performance of the baseline and KD models. Therefore, our baseline model performance is slightly better than that reported in~\cite{gong_psla,gong21b_interspeech}. We also intentionally minimize the training setting differences (e.g., mixup and augmentations) between the CNN and AST models to ensure performance differences are due to the architecture, rather than the training settings.

\section{FSD50K Experiments}
\label{sec:fsd50k}

\subsection{Which model is a good teacher?}
\label{sec:teacherstudent}

We first explore the optimal teacher model for each student model. Specifically, we use the KD setting mentioned in Section~\ref{sec:method} and test a set of \texttt{(teacher, student)} model pairs. Specifically, we consider the following models: EfficientNet B0, B2, B6~\cite{tan2019efficientnet} and DenseNet-121~\cite{8099726} (CNN models); and AST Tiny, Small, Base~\cite{gong21b_interspeech} (Transformer models). The reason for selecting these models are: 1) EfficientNet-B2 and AST-Base are the best CNN and Transformer models, and we want to see if KD can further improve their performance; 2) AST-Base is larger than EfficientNet-B2 in terms of the number of parameters. We include models with various sizes to clearly identify whether the improvement of KD is due to different model sizes or different model architecture; 3) We also include one different CNN architecture (DenseNet) to check if the cross-model KD framework generalizes to different CNN architectures. 

For each \texttt{(teacher, student)} pair, we select the temperature $\tau\in\{1.0, 2.5\}$ based on the performance on the validation set, all other knowledge distillation settings are identical for all \texttt{(teacher, student)} pairs. The mean average precisions (mAPs) of the student model of various \texttt{(teacher,student)} pairs on the FSD50K evaluation set are shown in Table~\ref{tab:teacherstudent}. Key observations are as follows:

\textbf{1. CNNs and ASTs are good teachers for each other.} While knowledge distillation improves the student model performance in almost all settings (denoted by $\dagger$ in Table~\ref{tab:teacherstudent}), we find that models always prefer a \emph{different} teacher. Specifically, all CNN models (including DenseNet) achieve the best performance when ASTs are the teacher; all AST models achieve the best performance when CNNs are the teacher. Note that even though DenseNet and EfficientNet models also have different architectures, knowledge distillation between these two models leads to limited improvement. This demonstrates that the inherent differences between CNNs and ASTs is crucial for the success of knowledge distillation. 


\textbf{2. For both directions, the student model matches or outperforms its teacher.} Usually, in knowledge distillation, the student model gets closer to, but is still weaker than, its teacher model. However, we find that for both \texttt{AST}$\rightarrow$\texttt{CNN} and \texttt{CNN}$\rightarrow$\texttt{AST}, the student model performs better than its teacher in most cases (denoted by $\ast$ in Table~\ref{tab:teacherstudent}), e.g., the best KD-CNN and KD-AST model achieves 58.2 and 61.7 mAP while their teacher model achieves only 56.3 and 55.7, respectively. This demonstrates that the student models are not simply copying the teacher model, but are able to combine the strength of the teacher with its own strength.

\begin{figure}[t]
  \centering
  \includegraphics[width=6.2cm]{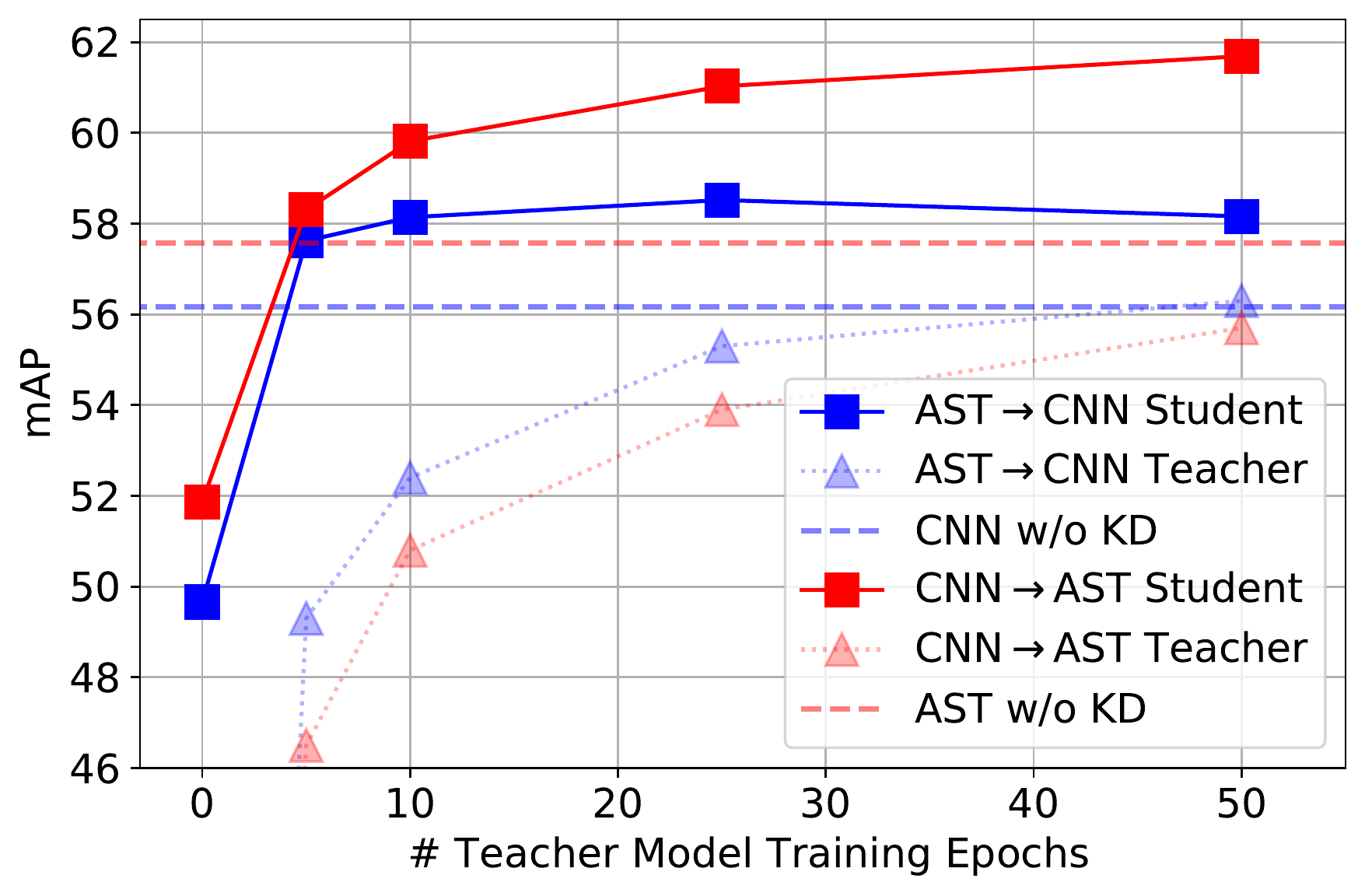}
  \caption{Performance correlation between teacher and student model when the model architectures are fixed. We save the checkpoint teacher models at epoch 0, 5, 10, 20, 30, 40, and 50 during teacher model training, then train student models with these checkpoint teacher models until convergence. We observe that for a fixed teacher model architecture, a stronger teacher trained with more epochs generally leads to a better student. Nevertheless, the student models outperform the no-KD baselines (the dashed lines) even when their teacher models are trained with only 5 epochs with poor performance. The examples we show in this figure are the optimal \texttt{(teacher,student)} pairs, i.e., \texttt{AST}$\rightarrow$\texttt{CNN}: \texttt{AST-Tiny}$\rightarrow$\texttt{EfficientNet-B2}; \texttt{CNN}$\rightarrow$\texttt{AST}: \texttt{EfficientNet-B0}$\rightarrow$\texttt{AST-Base}.}
  \label{fig:teacher_convergence}
\end{figure}

\textbf{3. The strongest teacher is not the best teacher.} Normally in knowledge distillation, the student model performance correlates with the teacher performance. However, we find for both \texttt{AST}$\rightarrow$\texttt{CNN} and \texttt{CNN}$\rightarrow$\texttt{AST}, the best teacher is not the strongest teacher. Instead, both CNN and AST perform better with a smaller (and weaker) teacher. We further verify that this is not because a stronger teacher requires a higher temperature, e.g., for \texttt{AST-Base} (strongest teacher) $\rightarrow$\texttt{EfficientNet-B2}, using a temperature of 2.5 (used in Table~\ref{tab:teacherstudent}) leads to an mAP of 57.3 while further increasing the temperature to 3.0 leads to a worse mAP of 57.1. For the same student model EfficientNet-B2, the smaller teacher AST-Tiny leads to a better mAP of 58.2. We conjecture it is because smaller models have similar properties (e.g., inductive bias) with larger models of the same type but are less overfit, and thus are better teachers. 

However, as shown in Figure~\ref{fig:teacher_convergence}, when the teacher model architecture is fixed, we find a stronger teacher trained with sufficient epochs generally leads to better student performance than a weaker teacher that is not adequately trained. Nevertheless, the student models still outperform the no-KD baselines even when their teacher models are trained with only 5 epochs with poor performance. This further confirms that the student is not simply copying the teacher's behavior as copying the behavior of a teacher weaker than the student itself can only cause a performance drop, but our experiments show that cross-model KD still leads to a performance improvement in this case.

\textbf{4. Self-KD leads to smaller or no improvement.} We observe that performance improvement can also sometimes be achieved by using the same teacher and student architecture (the main diagonal of Table~\ref{tab:teacherstudent}), particularly for AST models. This is as expected because 1) the soft labels helps optimization of the student model, and 2) since we use strong data augmentation (e.g., SpecAugment, mixup, random noise, and time shift) during training, knowledge distillation with the same model architecture can have a similar effect to self-training~\cite{xie2020self}. Nevertheless, the improvement led by self-knowledge distillation is much smaller than cross-model knowledge distillation, which demonstrates the importance of using a different teacher for knowledge distillation. 

\textbf{5. Iterative knowledge distillation does not further improve model performance.}
Since cross-model knowledge distillation works bi-directionally, the performance of both KD-CNN and KD-AST are better than no-KD baselines. Will using KD models instead of no-KD models as teachers lead to even better students? The answer is no, as shown in Table~\ref{tab:iter_kd}. While KD models (with their optimal teachers) are stronger, they are worse teachers than the no-KD models. This again proves our finding that a stronger teacher is not necessarily a better teacher as the student is not simply copying the teacher's behavior.

To summarize, in this section, we find both CNN and AST models prefer a \emph{small} and \emph{different} teacher. Cross-model knowledge distillation is beyond simple behavior copying but more about transferring and combining the strength of the teacher and student models as the performance of the teacher is surprisingly not important. 

\begin{table}[t]
\setlength\tabcolsep{2.2pt}
\centering
\caption{Performance of iterative knowledge disitllation. Though the KD models are stronger than no-KD models, they are worse teachers. Therefore, iterative knowledge distillation cannot further improve the performance.}
\begin{tabular}{@{}lccc@{}}
\toprule
\multicolumn{1}{c}{\begin{tabular}[c]{@{}c@{}}Student\\ Model\end{tabular}} & \begin{tabular}[c]{@{}c@{}}Teacher\\ Model\end{tabular} & \begin{tabular}[c]{@{}c@{}}Teacher\\ Performance\end{tabular} & \begin{tabular}[c]{@{}c@{}}Student\\ Performance\end{tabular} \\ \midrule
\multicolumn{4}{l}{{\color[HTML]{656565} No KD}}                                                                                                                                                                                                                      \\
EfficientNet-B2                                                             & -                                                       & -                                                             & 56.2                                                          \\
AST-Base                                                                    & -                                                       & -                                                             & 57.6                                                          \\ \midrule
\multicolumn{4}{l}{{\color[HTML]{656565} KD Iteration 1}}                                                                                                                                                                                                             \\
EfficientNet-B2                                                             & AST-Tiny w/o KD                                         & 56.3                                                          & 58.2                                                          \\
AST-Base                                                                    & EfficientNet-B0 w/o KD                                  & 55.7                                                          & 61.7                                                          \\ \midrule
\multicolumn{4}{l}{{\color[HTML]{656565} KD Iteration 2}}                                                                                                                                                                                                             \\
EfficientNet-B2                                                             & AST-Tiny w/ KD                                          & 58.6                                                          & 58.0                                                          \\
AST-Base                                                                    & EfficientNet-B0 w/ KD                                   & 57.4                                                          & 61.4                                                          \\ \bottomrule
\end{tabular}
\label{tab:iter_kd}
\end{table}

\subsection{Impact of Knowledge Distillation Settings}
\label{sec:kd_setting}

\begin{table}[t]
\centering
\caption{Student model performance on the FSD50K evaluation set due to various knowledge distillation settings. We conduct this experiment with the optimal teacher student pairs, i.e., \texttt{AST}$\rightarrow$\texttt{CNN}: \texttt{AST-Tiny}$\rightarrow$\texttt{EfficientNet-B2}; \\ \texttt{CNN}$\rightarrow$\texttt{AST}: \texttt{EfficientNet-B0}$\rightarrow$\texttt{AST-Base}.}
\begin{tabular}{@{}lcc@{}}
\toprule
KD Setting                   & \multicolumn{1}{l}{\texttt{AST}$\rightarrow$\texttt{CNN}} & \multicolumn{1}{l}{\texttt{CNN}$\rightarrow$\texttt{AST}} \\ \midrule
No KD Baseline               & 56.2                                          & 57.6                                          \\ \midrule
\multicolumn{3}{l}{{\color[HTML]{656565} KD Temperature $\tau$}}                                                                \\
$\tau$ = 0.5                    & 56.7                                          & 60.9                                          \\
$\tau$ = 1.0                    & 57.2                                          & \textbf{61.7}                                 \\
$\tau$ = 2.5                    & \textbf{58.2}                                 & 61.2                                          \\ \midrule
\multicolumn{3}{l}{{\color[HTML]{656565} KD Balancing Coefficient $\lambda$}}                                                               \\
$\lambda$ = 0 (No ground truth label)     & 56.4                                          & 59.7                                          \\
$\lambda$ = 0.25             & 58.0                                          & 60.9                                          \\
$\lambda$ = 0.5 (Default)    & \textbf{58.2}                                  & \textbf{61.7}                                          \\
$\lambda$ = 0.75             & 57.2                                          & 61.3                                          \\
$\lambda$ = 1.0 (No KD)      & 56.2                                          & 57.6                                          \\ \midrule
\multicolumn{3}{l}{{\color[HTML]{656565} Teacher Type}}                                                                      \\
Conventional Teacher         & 58.2                                          & 60.4                                          \\
Consistent Teacher (Default) & \textbf{58.2}                                 & \textbf{61.7}                                 \\ \midrule
\multicolumn{3}{l}{{\color[HTML]{656565} Separate Distillation Head}}                                                            \\
Yes                       & 56.9                                          & 61.4                                          \\
No (Default)             & \textbf{58.2}                                 & \textbf{61.7}                                 \\ \bottomrule
\end{tabular}
\label{tab:kd_setting}
\end{table}

We next study the performance impact of the knowledge distillation setting. Specifically, we focus on the optimal \texttt{(teacher,student)} pair found in the previous section, i.e., \texttt{AST-Tiny}$\rightarrow$\texttt{EfficientNet-B2} and \texttt{EfficientNet-B0}$\rightarrow$\texttt{AST-Base}, and set the knowledge distillation setting mentioned in Section~\ref{sec:kd} (marked as default in Table~\ref{tab:kd_setting}) as the base setting, and then change one factor at a time to observe the performance change. We show the results in Table~\ref{tab:kd_setting}. 

\textbf{KD Temperature $\tau$:} In Section~\ref{sec:teacherstudent}, we vary the temperature $\tau$ from \{1.0, 2.5\}. We find a higher temperature of 2.5 is better for \texttt{AST}$\rightarrow$\texttt{CNN} and a lower temperature of 1.0 is better for all other \texttt{(teacher,student)} pairs. Using an even lower temperature (or ``hard'' knowledge distillation~\cite{touvron2021training} used in the vision domain) cannot lead to good performance. This is due to audio classification usually being a multi-label classification task (each audio clip usually has more than one label), which is different from the single-label image classification task.

\textbf{KD Balancing Coefficient $\lambda$:} As shown in Table~\ref{tab:kd_setting}, balancing the ground truth label and teacher's soft prediction with a factor of $\lambda=0.5$ leads to the optimal performance for both \texttt{AST}$\rightarrow$\texttt{CNN} and \texttt{CNN}$\rightarrow$\texttt{AST}. Merely using the soft prediction of the teacher model can still improve the model, but with a much smaller performance gain. 

\textbf{Consistent teacher makes a difference:} We compare the student model performance with a consistent teacher~\cite{beyer2021knowledge} and a conventional teacher~\cite{hinton2015distilling}. The consistent teacher generates predictions based on the same augmented input with the student, while the conventional teacher generates prediction based on the original, unaugmented input. We find that in the \texttt{CNN}$\rightarrow$\texttt{AST} direction, a consistent teacher leads to a noticeable 1.3 mAP improvement while for \texttt{AST}$\rightarrow$\texttt{CNN}, it does not hurt performance. Therefore, consistent teaching should be used as a default setting.

\textbf{Separate Knowledge Distillation Head:} In the vision domain, knowledge distillation with a separate distillation head~\cite{touvron2021training} was shown to have better performance than conventional knowledge distillation~\cite{hinton2015distilling}. We compare these two settings for the audio classification task. Specifically, for a KD model with a separate distillation head, for \texttt{CNN}$\rightarrow$\texttt{AST}, we follow~\cite{touvron2021training} to append a distillation token~\texttt{[DIS]} to the AST token sequence and use a separate linear head to map the output of \texttt{[DIS]} token to the label space. During knowledge distillation, the output of the original~\texttt{[CLS]} token is used to generate the loss versus the ground truth label $Loss_g$ while the output of the new~\texttt{[DIS]} token is used to generate the loss versus the teacher prediction $Loss_d$. Similarly, for \texttt{AST}$\rightarrow$\texttt{CNN}, we use two independent linear heads to map the representation of the penultimate layer to the label space, we then use the output of one head to generate $Loss_g$ and another to generate $Loss_d$.
As shown in Table~\ref{tab:kd_setting}, while the more complex separate distillation head solution still outperforms the baseline model, the performance gains are smaller for both \texttt{AST}$\rightarrow$\texttt{CNN} and \texttt{CNN}$\rightarrow$\texttt{AST} for audio classification, which is different from the finding in the vision domain. We, therefore, stick to the conventional knowledge distillation~\cite{hinton2015distilling} and do not use a separate distillation head.

To summarize, throughout Table~\ref{tab:kd_setting}, we find that knowledge distillation with any setting improves model performance. But the optimal setting is quite different from that of the vision domain~\cite{touvron2021training}. In general, multi-label audio classification knowledge distillation requires an appropriate temperature larger than 1, a balance between the ground truth label and teacher model prediction, and consistent teaching~\cite{beyer2021knowledge}. It is sufficient to just follow conventional knowledge distillation to generate both $Loss_g$ and $Loss_d$ using the original student model prediction without adding a separate distillation head.

\begin{figure}[t]
    \centering
    \subfloat{{\includegraphics[width=4.5cm]{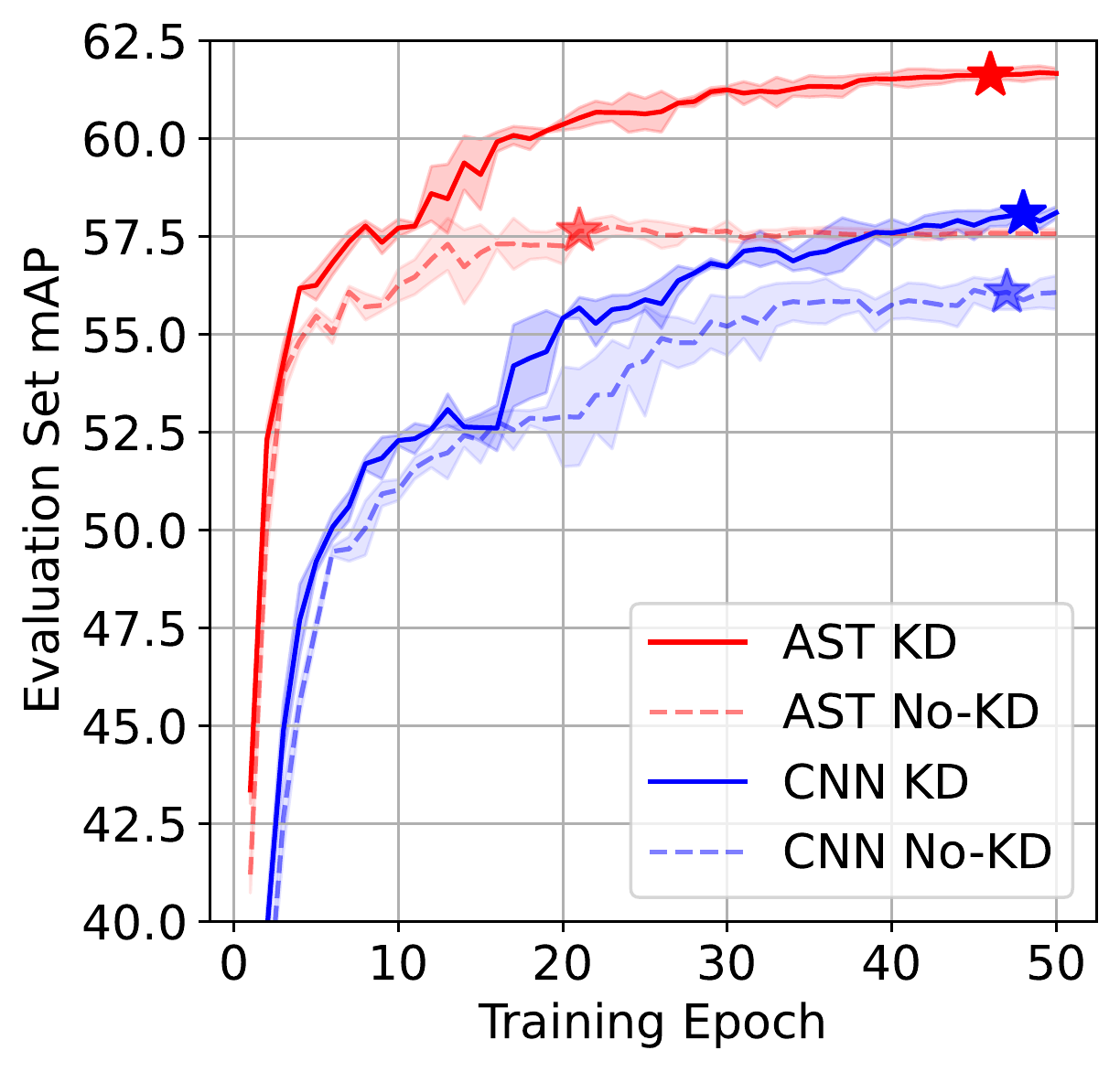} }}%
    \subfloat{{\includegraphics[width=4.5cm]{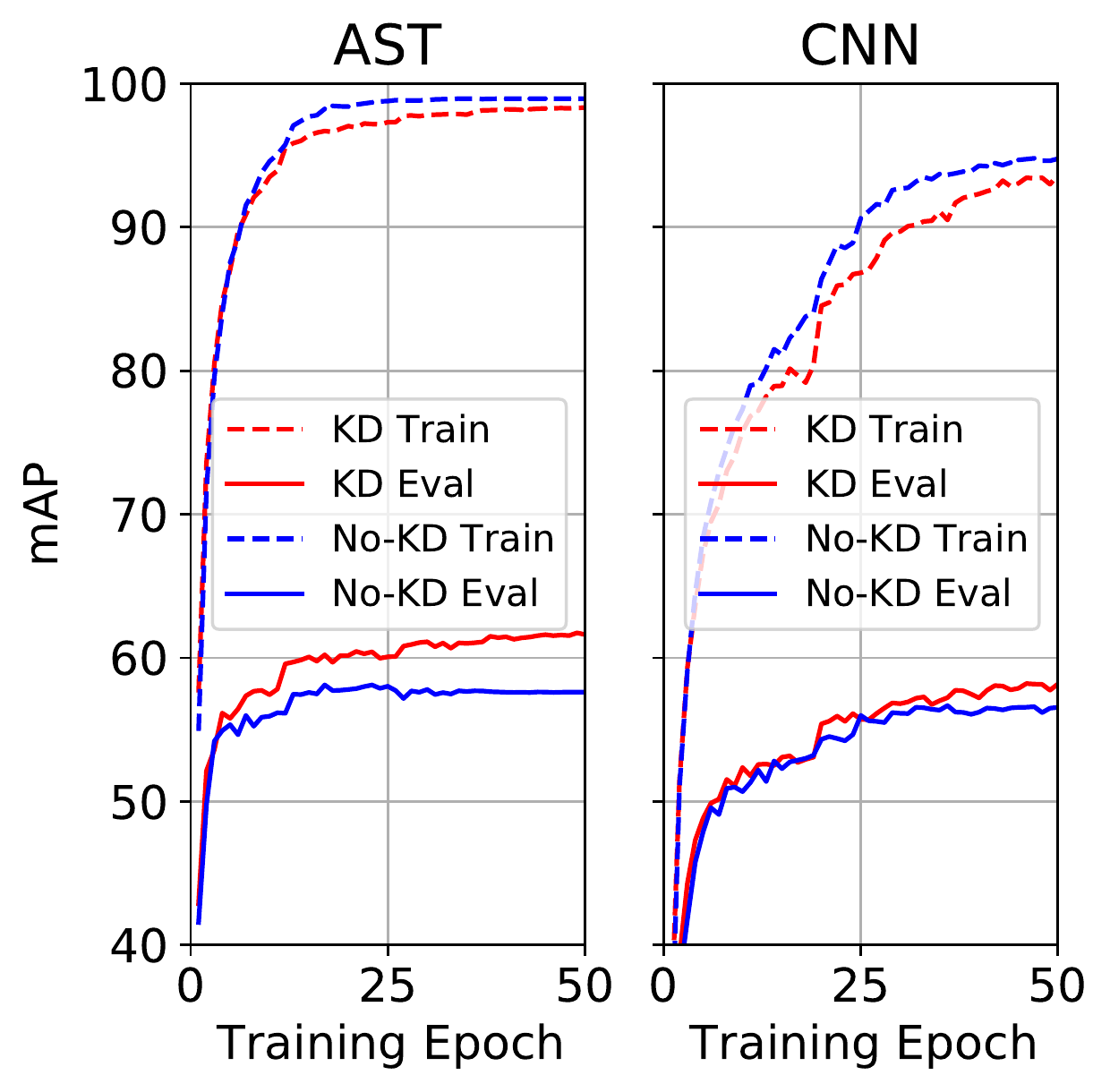} }}%
    \caption{\underline{Left}: Learning curve of the CNN model (EfficientNet-B2) and AST model (AST-Base) trained with and without KD on FSD50K. $\star$ denotes the epoch that the model achieves the best performance on the validation set. Each experiment is run three times, and the standard deviation is shown in the shade. The learning rate is cut in half when the model performance on the validation set does not improve for 2 epochs so the three runs of each experiment have slightly different learning rate schedulers and the performance variance is relatively large around the epochs of the learning rate change. \underline{Right}: Training and evaluation mAP of AST and CNN models with and without KD. KD models have a higher evaluation mAP but lower training mAP, indicating they are less overfit.}
    \label{fig:lr_curve}
\end{figure}

\begin{table*}[t]
\centering
\caption{Prediction similarity in Pearson correlation coefficient (upper part) and model representation similarity in singular vector canonical correlation analysis~\cite{raghu2017svcca} (lower part) of various model pairs. Similarity value between KD model with corresponding teacher is underlined.}
\begin{tabular}{@{}lcccccc@{}}
\toprule
\multicolumn{1}{c}{\textit{}} & \begin{tabular}[c]{@{}c@{}}EfficientNet-B0\\ w/o KD\end{tabular} & \begin{tabular}[c]{@{}c@{}}EfficientNet-B2\\ w/o KD\end{tabular} & \begin{tabular}[c]{@{}c@{}}EfficientNet-B2\\ w/ KD\end{tabular} & \begin{tabular}[c]{@{}c@{}}AST-Tiny\\ w/o KD\end{tabular} & \begin{tabular}[c]{@{}c@{}}AST-Base\\ w/o KD\end{tabular} & \begin{tabular}[c]{@{}c@{}}AST-Base\\ w/ KD\end{tabular} \\ \midrule
\multicolumn{7}{l}{\textit{Prediction Correlation}}                                                                                                                                                                                                                                                                                                                                                                      \\
EfficientNet-B2 w/o KD        & 0.861                                                            & 1.000                                                            & 0.870                                                           & 0.804                                                     & 0.791                                                     & 0.865                                                    \\
EfficientNet-B2 w/ KD         & 0.862                                                            & 0.870                                                            & 1.000                                                           & \underline{0.833}                                                     & 0.810                                                     & 0.882                                                    \\
AST-Base w/o KD               & 0.786                                                            & 0.791                                                            & 0.810                                                           & 0.844                                                     & 1.000                                                     & 0.880                                                    \\
AST-Base w/ KD                & \underline{0.880}                                                            & 0.865                                                            & 0.882                                                           & 0.880                                                     & 0.880                                                     & 1.000                                                    \\ \midrule
\multicolumn{7}{l}{\textit{SVCCA}}                                                                                                                                                                                                                                                                                                                                                                                       \\
EfficientNet-B2 w/o KD        & 0.720                                                            & 1.000                                                            & 0.734                                                           & 0.705                                                     & 0.651                                                     & 0.706                                                    \\
EfficientNet-B2 w/ KD         & 0.724                                                            & 0.734                                                            & 1.000                                                           & \underline{0.718}                                         & 0.695                                                     & 0.721                                                    \\
AST-Base w/o KD               & 0.651                                                            & 0.651                                                            & 0.695                                                           & 0.779                                                     & 1.000                                                     & 0.784                                                    \\
AST-Base w/ KD                & \underline{0.746}                                                            & 0.706                                                            & 0.721                                                           & 0.807                                                     & 0.784                                                     & 1.000                                                    \\ \bottomrule
\end{tabular}
\label{tab:similarity}
\end{table*}

\subsection{How Does CMKD Improve Performance?}
\label{sec:analysis}

In the previous two sections, we see that knowledge distillation between CNN and AST models noticeably boosts model performance. It is natural to ask why it works so well. In this section, we try to answer this question with a series of probing tests and analysis. We focus on the optimal \texttt{(teacher,student)} pair found in Section~\ref{sec:teacherstudent}, i.e., \texttt{AST-Tiny}$\rightarrow$\texttt{EfficientNet-B2} for \texttt{AST}$\rightarrow$\texttt{CNN} and \texttt{EfficientNet-B0}$\rightarrow$\texttt{AST-Base} for \texttt{CNN}$\rightarrow$\texttt{AST}, respectively. Key findings are as follows: 

\textbf{Knowledge distillation makes the model learn faster and be less prone to overfitting}.
We compare the learning curve of the CNN model (EfficientNet-B2) and the AST model (AST-Base) trained with and without knowledge in Figure~\ref{fig:lr_curve} (left). We find that for both CNN and AST models, knowledge distillation helps the models learn faster and be less prone to overfitting. The KD models outperform the no-KD models from the first epoch, and consistently improves until the end of training while the performance of no-KD models saturate much earlier. For example, at the epoch that the no-KD AST achieves the best performance (the $21^{st}$ epoch), KD AST already outperforms no-KD AST by 2.9 mAP with its faster learning; from the $21^{st}$ epoch to the last epoch, KD AST further enlarges its advantage over no-KD AST from 2.9 mAP to 4.1 mAP. In Figure~\ref{fig:lr_curve} (right), we show the training and evaluation mAP of AST and CNN models with and without KD. Both KD-AST and KD-CNN models have a higher evaluation mAP but lower training mAP, confirming they are less overfit.

\textbf{Knowledge distillation narrows the gap between two classes of models.} Next, we analyze the similarity between CNN and AST models before and after cross-model knowledge distillation. Specifically, we show the prediction similarity in Pearson correlation coefficient (PCC) and model representation similarity in singular vector canonical correlation analysis (SVCCA)~\cite{raghu2017svcca}. Since FSD50K is a multi-label classification task, we average the PCC of the predictions of each evaluation sample of the two models being compared as the prediction similarity value. SVCCA is a technique to measure the similarity of two representations. It has nice properties such as affine transform invariance and supports the comparison of two representations of two different dimensions. We use SVCCA with the singular vector that explains 30\% variance of each representation as the model representation similarity. 

As shown in Table~\ref{tab:similarity}, we find that cross-model distillation not only makes the student model more similar to its teacher, but also to models of the same class with its teacher. For example, for \texttt{EfficientNet-B0}$\rightarrow$\texttt{AST-Base}, after knowledge distillation, not only the prediction similarity score between AST-Base (student) and Efficient-B0 (teacher) increases from 0.786 to 0.880, but also the similarity score between AST-Base (student) and EfficientNet-B2 (another CNN model with different architecture and independent training process) increases from 0.791 to 0.865. This phenomenon is consistent for both CNN and AST models, and for both prediction and model representation similarity, indicating that cross-model knowledge distillation not only makes the specific teacher and student model instances similar, but actually narrows the gap between two classes of models of CNN and AST.

\begin{figure}[t]
  \centering
  \includegraphics[width=8.5cm]{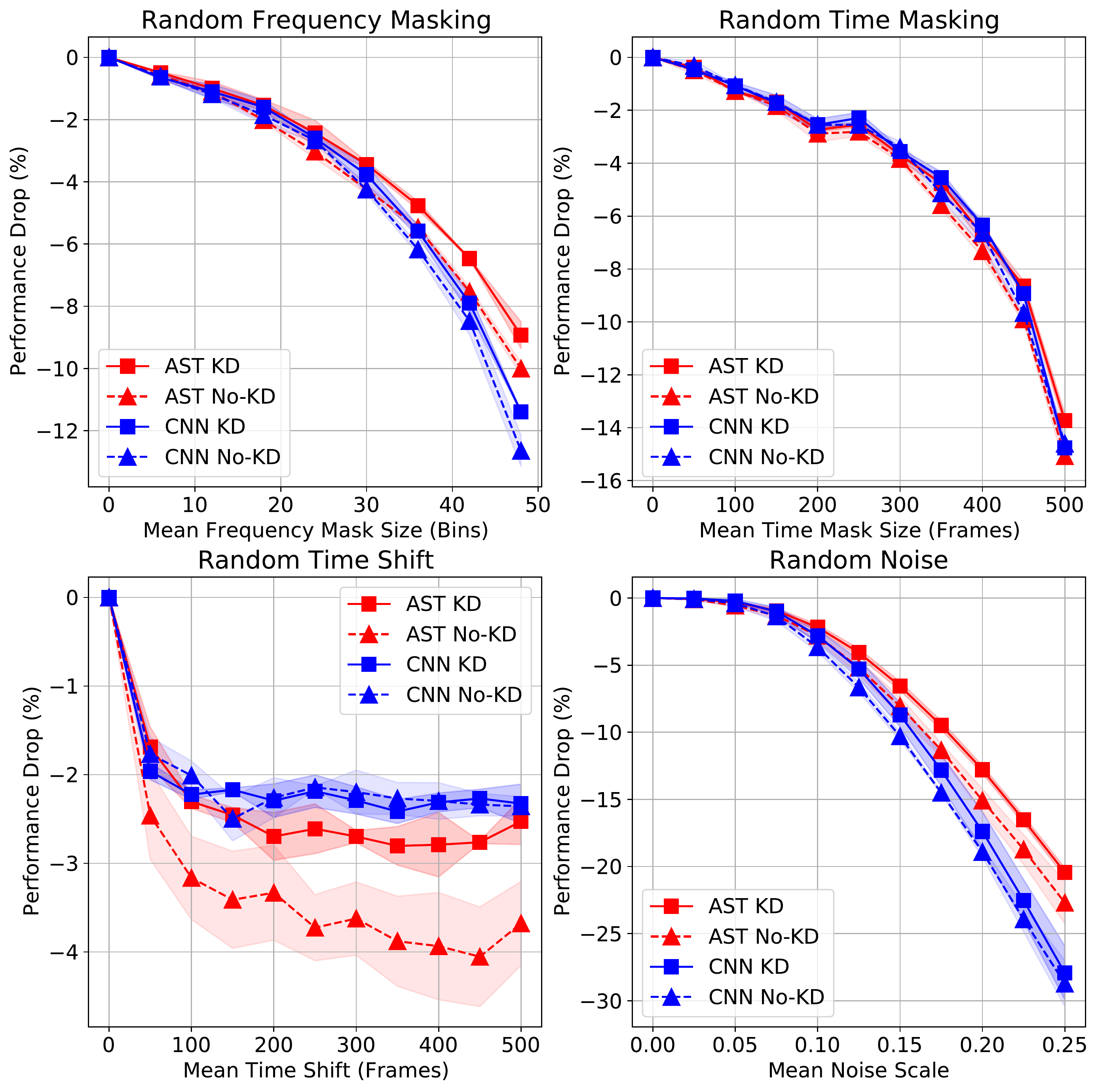}
  \caption{Model performance drop (\%) on input with various type and level of input contamination during inference. Upper left: random frequency masking, upper right: randome time masking; lower left: random time shift; lower right: random noise. Each experiment is run three times, and the stand deviation is shown in the shade.}
  \label{fig:robust}
\end{figure}

\begin{figure}[t]
  \centering
  \includegraphics[width=6cm]{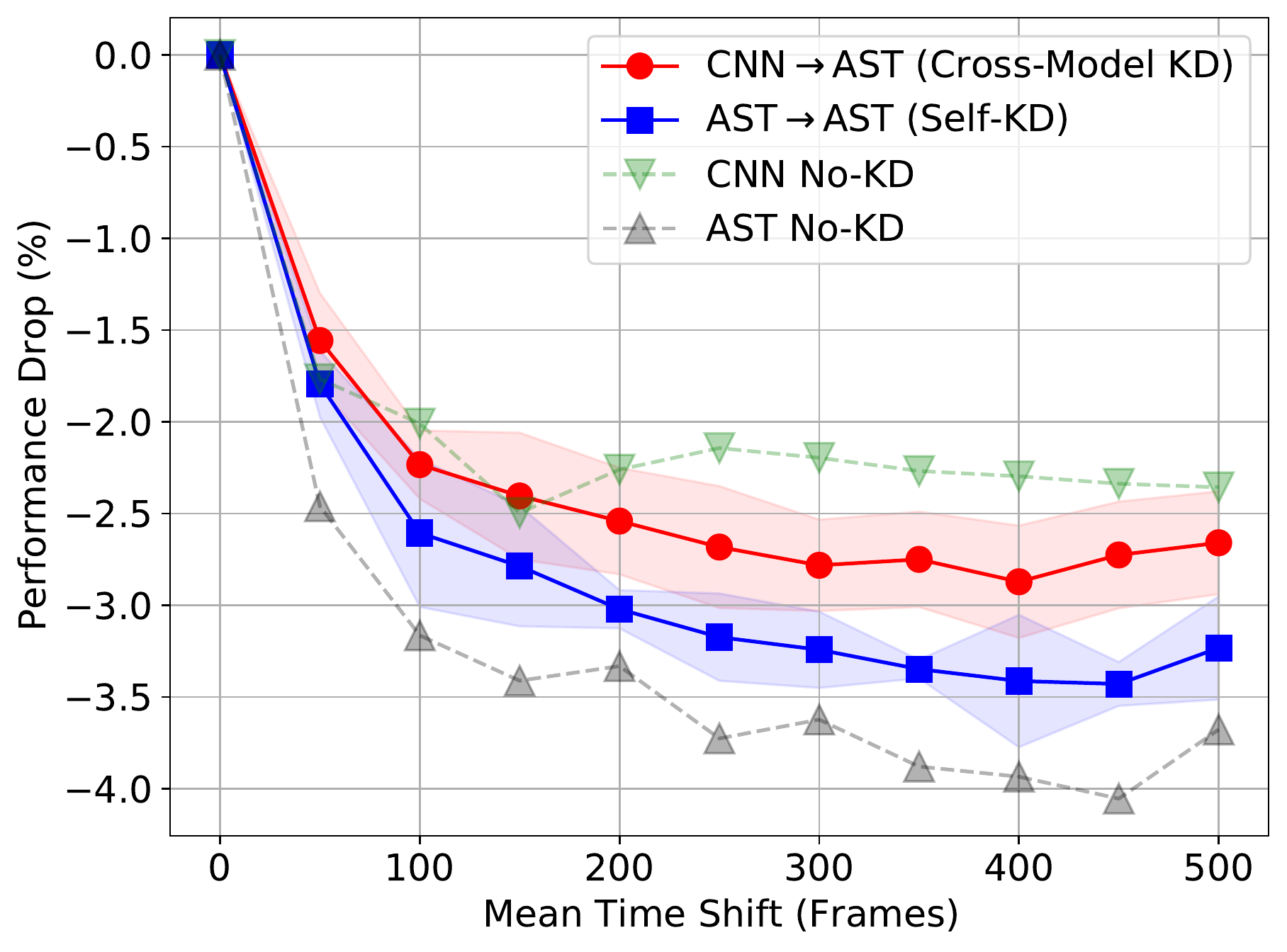}
  \caption{Comparison of the robustness of AST models trained with self-KD and CMKD to input time shifting. We show the mean and standard deviation of the performance drop of AST models trained with various CNN teachers (EfficientNet-B0, B2, B6, and DenseNet 121) in red (CMKD), and those of AST models trained with various AST models (AST-Tiny, Small, and Base) in blue (self-KD). AST models trained with CMKD are more time shift-invariant than those trained with self-KD, indicating CMKD indeed transfers the translation invariance property from CNN to AST.}
  \label{fig:robust2}
\end{figure}

\textbf{Knowledge distillation improves model robustness.} As mentioned in Section~\ref{sec:diff}, the major properties of CNN models are \emph{translation invariance} and \emph{locality} while one major property of AST is \emph{non-locality} as it looks at all of the context of earlier layers. Therefore, CNN models are more robust to input shifting but are more prone to local feature contamination while AST models are the opposite~\cite{naseer2021intriguing}. We verify this point by evaluating the models on a contaminated FSD50K evaluation set. As shown in Figure~\ref{fig:robust}, when knowledge distillation is not applied, the CNN model is indeed less impacted by random time-shifting (input shifting) while the AST model is less impacted by random frequency masking and random noise (local feature contamination). 

Interestingly, we find cross-model knowledge distillation not only improves models' mean average precision on clean input, but also enhances their robustness to contaminated input, e.g., KD-AST inherits its CNN teacher's advantage of translation invariance and is significantly more robust to random time shifts. Similarly, the CNN model inherits its AST teacher's property and is more robust to frequency masking and random noise. It is worth noting that random time-shifting and noise are used as training data augmentation for all models, which already encourages the models to be robust. Therefore, the robustness improvement led by cross-model KD might be underestimated. 

Are such robustness improvements led by the knowledge transferred from the teacher, or are they a product of the knowledge distillation process itself (e.g., the effect of the soft label and consistent teaching)? To answer this question, we further compare the robustness to random time-shifting of AST models trained with cross-model KD and self-KD. For self-KD AST, we evaluate the AST model trained with AST-Tiny, AST-Small, and AST-Base teachers. For cross-model KD, we evaluate AST model trained with EfficientNet-B0, EfficientNet-B2, EfficientNet-B6, and DenseNet-121 teachers. As shown in Figure~\ref{fig:robust2}, both cross-model KD and self-KD lead to a robustness improvement, but cross-model KD is noticeably more effective, demonstrating both the knowledge transferred from the teacher and the KD process contributes to robustness improvements. Our finding is consistent with previous work~\cite{abnar2020transferring} that showed knowledge distillation can transfer model properties such as inductive bias and allow models to get advantages from each other, which is one major reason for the success of cross-model knowledge distillation. 

\begin{figure}[h]
  \centering
  \includegraphics[width=5.5cm]{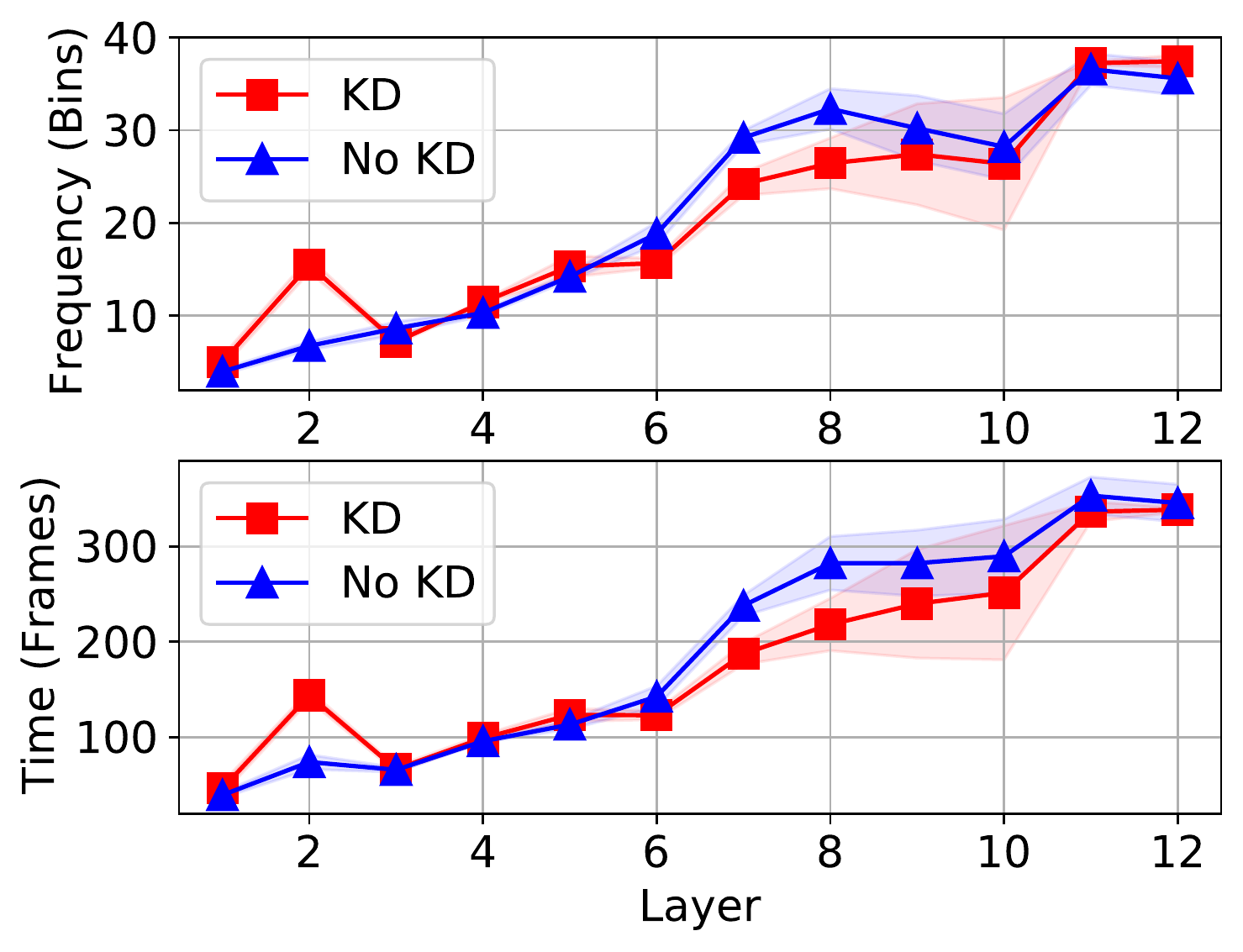}
  \caption{AST model's mean attention distance of each layer in frequency (upper) and time (lower) dimension before and after KD. KD noticeably changes the attention behavior of the $2^{nd}$ layer and $7-10^{nd}$ layers.}
  \label{fig:att_dis}
\end{figure}

\textbf{Knowledge distillation changes the attention map of AST.} Finally, we observe an interesting phenomenon that KD noticeably changes the attention map of the AST model. As discussed in Section~\ref{sec:diff}, the AST model's receptive field is learnable and can be arbitrarily large while the receptive field of the CNN model is manually set and fixed. In Figure~\ref{fig:att_dis}, we visualize the AST model's mean attention distance of each layer in the frequency and time dimension. The attention distance is the distance between a given query token and all other tokens averaged by the attention weight between that query token and each other token, which is analogous to the receptive field. As expected, we observe that the higher layers of AST have a larger mean attention distance, indicating AST can learn hierarchical features from the data. Interestingly, we find KD changes the attention behavior of AST. Particularly, the attention distance of the second layer almost doubled, and the $7^{th}$-$10^{th}$ layers slightly decrease. This means the KD-AST model captures more global information at lower layers, which potentially is one key reason for the performance improvement.

\subsection{Multiple-Teacher Knowledge Distillation}

In Sections~\ref{sec:teacherstudent}-\ref{sec:analysis}, we use a \emph{single} teacher model during knowledge distillation. In this section, we further explore whether using \emph{multiple} teachers~\cite{ren2021co} improves performance. Specifically, we consider two types of multi-teacher knowledge distillation: \emph{1) Multi-Loss KD}: the student model is trained with multiple distillation losses, each corresponds to a teacher model. Each distillation loss is balanced with the coefficient of $(1-\lambda)/n$, where $n$ is the number of teachers. \emph{2) Ensemble KD}: The predictions of the teachers are averaged and a single distillation loss is applied. For both CNN and AST students, we test a set of different combinations of teachers. We mainly consider teacher committees that consist of models of a different class with the student model based on our finding in Section~\ref{sec:teacherstudent}. We also try using both AST and CNN as the teachers.

As shown in Table~\ref{tab:mul_teacher}, multi-teacher knowledge distillation further improves the performance of the AST model. While any teacher combination and multi-teacher KD setting leads to better performance than single-teacher KD, we again observe the importance of diversity in the teacher models. When the teacher committee consists of different model architectures (EfficientNet and DenseNet) and the teacher models' type is different from the student model's type (CNN and AST), KD leads to the best performance. Multi-teacher KD, however, does not work for \texttt{AST}$\rightarrow$\texttt{CNN}.

\begin{table}[t]
\centering
\caption{Results with multi-teacher knowledge distillation. We consider various combinations of teachers and two types of multi-teacher knowledge distillation: \emph{1) Multi-Loss KD}: the student model is trained with multiple distillation losses, each corresponds to a teacher model and is balanced with the coefficient of $(1-\lambda)/n$, where $n$ is the number of teachers. \emph{2) Ensemble KD}: The predictions of the teachers are averaged and a single distillation loss is applied.}
\begin{tabular}{@{}llcc@{}}
\toprule
Student Model   & \multicolumn{1}{c}{Teacher Model}                                           & Multi-Loss   & Ensemble  \\ \midrule
EfficientNet-B2 & \quad\quad AST-Tiny                                                                    & \multicolumn{2}{c}{\textbf{58.2}} \\ \cmidrule(lr){2-2}
EfficientNet-B2 & \begin{tabular}[c]{@{}l@{}} \quad\quad AST-Tiny\\ \quad + AST-Small\end{tabular}              & 58.0         & 57.9      \\ \cmidrule(lr){2-2}
EfficientNet-B2 & \begin{tabular}[c]{@{}l@{}}\quad\quad AST-Tiny\\ \quad + AST-Small\\ \quad + AST-Base\end{tabular} & 57.4         & 58.2      \\ \cmidrule(lr){2-2}
EfficientNet-B2 & \begin{tabular}[c]{@{}l@{}}\quad\quad AST-Tiny \\ + EfficientNet-B0\end{tabular}        & 57.7         & 58.1      \\ \midrule
AST-Base        & \ \  EfficientNet-B0                                                             & \multicolumn{2}{c}{61.7} \\ \cmidrule(lr){2-2}
AST-Base        & \begin{tabular}[c]{@{}l@{}}\ \  EfficientNet-B0\\ + EfficientNet-B2\end{tabular} & 61.8         & 62.3      \\ \cmidrule(lr){2-2}
AST-Base        & \begin{tabular}[c]{@{}l@{}}\ \  EfficientNet-B0 \\ + DenseNet-121\end{tabular}   & \textbf{62.3}         & 62.2      \\ \cmidrule(lr){2-2}
AST-Base        & \begin{tabular}[c]{@{}l@{}}\ \  EfficientNet-B0\\ \quad + AST-Tiny\end{tabular}        & 62.1         & 61.9      \\ \bottomrule
\end{tabular}
\label{tab:mul_teacher}
\end{table}

\subsection{KD with CNN-Attention Hybrid Model}
\label{sec:hybrid}

In previous sections, we focus on cross-model KD between \emph{pure} CNN and Transformer models. Recently, a few \emph{CNN-attention hybrid models} have been proposed that combine CNN and attention models by integrating their model components~\cite{srivastava2021conformer,gong_psla,ford2019deep,miyazaki2020convolution,kong2020sound}. How does cross-model KD work with such CNN-Attention hybrid models? In this section, we consider a CNN-Attention hybrid model proposed in~\cite{gong_psla}. The hybrid model uses EfficientNet as the feature extractor and adds a 4-headed self-attention attention module on top of the EfficientNet. Without knowledge distillation, the hybrid model performs slightly better than the pure CNN model on FSD50K. We apply the same default KD setting and study knowledge distillation of \texttt{Hybrid}$\rightarrow$\texttt{CNN} and \texttt{Hybrid}$\rightarrow$\texttt{AST} (hybrid model as teacher) as well as \texttt{CNN}$\rightarrow$\texttt{Hybrid} and \texttt{AST}$\rightarrow$\texttt{Hybrid} (hybrid model as student). For the hybrid model, we consider a small model hybrid-B0 and a large model hybrid-B2 with EfficientNet-B0 and EffcientNet-B2 feature extractors, respectively.

\begin{table}[t]
\setlength\tabcolsep{4.5pt}
\caption{Results of CNN-attention hybrid model knowledge distillation on FSD50K. CNN-Attention hybrid model is good teacher of both pure CNN and AST models. Conversely, the performance of CNN-Attention hybrid model can also be noticeably improved when trained with KD with both pure CNN and AST models.}
\centering
\begin{tabular}{@{}lcc@{}}
\toprule
\multicolumn{1}{c}{\begin{tabular}[c]{@{}c@{}}KD Setting \\ (Teacher $\rightarrow$ Student)\end{tabular}} & \begin{tabular}[c]{@{}c@{}}Teacher\\ Performance\end{tabular} & \begin{tabular}[c]{@{}c@{}}Student Performance\\ (No KD$\rightarrow$KD)\end{tabular} \\ \midrule
Hybrid-B0 $\rightarrow$ EfficientNet-B2    & 56.3                                                          & 56.2$\rightarrow$57.9                                                                  \\
Hybrid-B0 $\rightarrow$ AST-Base           & 56.3                                                          & 57.6$\rightarrow$61.9                                                                  \\
Hybrid-B2 $\rightarrow$ EfficientNet-B2           & 56.2                                                          & 56.2$\rightarrow$57.4                                                                 \\
Hybrid-B2 $\rightarrow$ AST-Base           & 56.2                                                          & 57.6$\rightarrow$61.7                                                                  \\\midrule
EfficientNet-B0 $\rightarrow$ Hybrid-B2    & 55.7                                                          & 56.2$\rightarrow$57.9                                                                  \\
AST-Tiny $\rightarrow$ Hybrid-B2           & 56.3                                                          & 56.2$\rightarrow$57.9                                                                  \\ \bottomrule
\end{tabular}
\label{tab:hybrid}
\end{table}

As shown in Table~\ref{tab:hybrid}, we find the CNN-attention hybrid model is a good teacher of both pure CNN and AST models. Specifically, \texttt{Hybrid}$\rightarrow$\texttt{AST} leads to a mAP of 61.9, which is slightly better than \texttt{CNN}$\rightarrow$\texttt{AST} (61.7 mAP); \texttt{Hybrid}$\rightarrow$\texttt{CNN} also achieves similar performance with \texttt{AST}$\rightarrow$\texttt{CNN}. Interestingly, we find the small hybrid model is also a better teacher than the large model, which is consistent with our finding in Section~\ref{sec:teacherstudent}. Conversely, the performance of the CNN-attention hybrid model can also be noticeably improved when trained with either a pure CNN or an AST teacher. Nevertheless, its performance is worse than KD-AST and KD-CNN.

To conclude, CNN-attention hybrid models are in the middle of pure CNN models and pure attention models. It is a good teacher of both CNN and AST, and conversely, its performance also boosts when trained with a CNN or AST teacher. Nevertheless, despite the more complex architecture, the hybrid model does not outperform either KD-CNN or KD-AST, demonstrating the advantage in simplicity and effectiveness of the proposed CNN-AST cross-model knowledge distillation. 

\subsection{Comparison with State-of-the-art}

Finally, we compare the best KD models with state-of-the-art models.
We summarize the performance improvement on the FSD50K evaluation set led by KD in Table~\ref{tab:fsd_best}.
As with previous work~\cite{gong_psla,gong21b_interspeech}, we find the best CNN model is EffcientNet-B2 and the best AST model is AST-Base on FSD50K. For EfficientNet-B2, we find its optimal teacher is AST-Tiny; for AST-Base, we find its optimal KD setting is using multiple CNN teachers (EfficientNet-B0 and DenseNet-121). In previous sections, we report the evaluation set mAP of the model checkpoint that achieves the highest mAP on the validation set. In addition to that, here we also report the performance of the model with weight averaging (WA)~\cite{izmailov2018averaging} and ensemble models. For weight averaging, we average all weights of the model checkpoints of multiple epochs, the start averaging epoch is selected using the validation set and the end averaging epoch is the last epoch of training (the $50^{th}$ epoch). Weight averaging does not increase the computation overhead of training and inference. For ensemble, we average the predictions of three models trained with the same setting but different random seeds. As with~\cite{gong_psla}, we find weight averaging and ensembles both improve the performance of all models. 

After applying weight averaging, the KD-CNN and KD-AST model achieves 58.5$\pm$0.2 and 62.9$\pm$0.1 mAP respectively, which both significantly outperform all existing models (summarized in Table~\ref{tab:fsd_comp}). After ensembling, the KD-CNN and KD-AST model achieves an even better performance of 60.7 and 63.4 mAP. 

\begin{table}[t]
\caption{Comparison of the mAP of CNN and AST models trained with and without KD in weight averaging (WA) and ensemble settings on FSD50K.}
\centering
\begin{tabular}{@{}lcccc@{}}
\toprule
\multicolumn{1}{c}{Student Model} & KD  & \begin{tabular}[c]{@{}c@{}}Best Model \\ on Validation Set\end{tabular} & w/ WA      & Ensemble \\ \midrule
EfficientNet-B2                   & No  & 56.2$\pm$0.5                                                              & 57.6$\pm$0.3 & 59.4     \\
EfficientNet-B2                   & Yes & 58.2$\pm$0.1                                                              & 58.5$\pm$0.2 & 60.7     \\
AST-Base                          & No  & 57.6$\pm$0.4                                                              & 60.1$\pm$0.2 & 61.1     \\
AST-Base                          & Yes & 62.3$\pm$0.1                                                              & 62.9$\pm$0.1 & 63.4     \\ \bottomrule
\end{tabular}
\label{tab:fsd_best}
\end{table}

\begin{table}[t]
\caption{Comparison with previous methods on FSD50K. Since previous methods do not use ensemble, we compare with them using our best \emph{single} model for fairness.}
\centering
\begin{tabular}{@{}lc@{}}
\toprule
\multicolumn{1}{c}{Model} & Mean Average Precision (mAP)       \\ \midrule
FSD50K Baseline~\cite{fonseca2020fsd50k}           & 43.4      \\
Wav2CLIP~\cite{wu2021wav2clip}                  & 43.1      \\
YAMNet~\cite{YAMNet}                    & 50.4      \\
Audio Transformers~\cite{verma2021audio}        & 53.7      \\
Shift-Invariant CNN~\cite{fonseca2021improving}       & 54.1      \\
PSLA~\cite{gong_psla}                      & 56.7      \\
AST~\cite{gong21b_interspeech}                       & 57.6      \\ \midrule
KD-CNN                    & 58.5$\pm$0.2 \\
KD-AST                    & 62.9$\pm$0.1 \\ \bottomrule
\end{tabular}
\label{tab:fsd_comp}
\end{table}

\section{AudioSet and ESC-50 Experiments}
\label{sec:other_dataset}

Throughout Section~\ref{sec:fsd50k}, we focus on FSD50K dataset due to the reasons mentioned in Section~\ref{sec:dataset}. In order to study the generalization of the proposed KD strategy, in this section, we further evaluate on AudioSet and ESC-50.

\subsection{AudioSet Experiments}
\label{sec:audioset}

AudioSet~\cite{gemmeke2017audio} is a collection of over 2 million 10-second audio clips excised from YouTube videos and labeled with the sounds that the clip contains from a set of 527 labels. The balanced training, full training, and evaluation set contain 22k, 2M, and 20k samples, respectively. To study the impact of training data size, we conduct experiments on both the balanced and full training set and evaluate on the AudioSet evaluation set. As the goal is to study the generalization of the proposed method, we do not search the KD hyperparameters with AudioSet but just re-use the optimal KD setting found with the FSD50K dataset in Table~\ref{tab:kd_setting}. We find small models such as EfficientNet-B0 and AST-Tiny do not perform well with AudioSet as the dimension of the penultimate layer representation is close to or smaller than the label dimension of AudioSet (527). Therefore, we focus on larger models EfficentNet-B2 and AST-Base. As with FSD50K, we report the performance in mAP of 1) model of the last training epoch, 2) model with weight averaging (WA), and 3) ensemble three models trained with same setting but different random seeds. We show the results in Table~\ref{tab:audioset}. The key findings are as follows:

\begin{table}[t]
\centering
\caption{Performance in mAP of CNN and AST models on the AudioSet evaluation set before and after knowledge distillation. $\diamond$ denotes the model is trained with extra 5 epochs (10 epochs in total).}
\begin{tabular}{@{}lcccc@{}}
\toprule
\multicolumn{1}{c}{Student Model}  & KD   & Last Epoch    & w/ WA  & Ensemble \\ \midrule
\multicolumn{5}{l}{{\color[HTML]{656565} \textit{Balanced AudioSet Training Set (20K)}}} \\
EfficientNet-B2                    & No   & 31.3$\pm$0.4  & 31.5$\pm$0.5            & 33.7     \\
EfficientNet-B2                    & Yes  & 34.4$\pm$0.2  & 34.5$\pm$0.2            & 35.6     \\
AST-Base                           & No   & 33.8$\pm$0.1  & 35.3$\pm$0.1            & 36.6     \\
AST-Base                           & Yes  & 36.2$\pm$0.2  & 36.4$\pm$0.2            & 37.6     \\ \midrule
\multicolumn{5}{l}{{\color[HTML]{656565} \textit{Full AudioSet Training Set (2M)}}}      \\
EfficientNet-B2                    & No   & 44.1$\pm$0.1  & 44.4$\pm$0.1            & 46.2     \\
EfficientNet-B2                    & Yes  & 45.3$\pm$0.1  & 45.4$\pm$0.1            & 46.2     \\
AST-Base                           & No   & 45.4$\pm$0.1  & 46.1$\pm$0.1            & 47.8     \\
AST-Base                           & Yes  & 46.8$\pm$0.1  & 46.8$\pm$0.1            & 48.1 \\
AST-Base$^\diamond$                & Yes  & 47.1$\pm$0.0  & 47.1$\pm$0.0            & 48.2 \\ \bottomrule
\end{tabular}
\label{tab:audioset}
\end{table}

\begin{table}[t]
\centering
\caption{Comparison with previous and concurrent methods on AudioSet. Since not all previous methods use ensemble and ensemble settings are usually different, we compare with them using our best single model for fairness. $\dagger$ denotes the experiment uses 32kHz audio, which can lead to about 0.4 mAP improvement~\cite{kong2020panns} compared with using 16kHz audio. $\star$ denotes concurrent efforts (\emph{arxiv} preprint). We use a more commonly used and computationally efficient sampling rate of 16kHz, but still, our KD-AST outperforms or matches all previous methods while our KD-CNN outperforms all previous CNN and CNN-attention models.}
\begin{tabular}{@{}lcc@{}}
\toprule
\multicolumn{1}{c}{Model}     & Type                               & mAP                         \\ \midrule
AudioSet Baseline~\cite{gemmeke2017audio}             & CNN                                & 31.4                        \\
TALNet~\cite{wang2019comparison}                        & CNN-Attention                      & 36.2                        \\
DeepRes~\cite{ford2019deep}                       & CNN-Attention                      & 39.2                        \\
PANNs~\cite{kong2020panns} $^\dagger$                         & CNN                                & 43.9                        \\
PSLA~\cite{gong_psla}                          & CNN-Attention                      & 44.4                        \\
AST~\cite{gong21b_interspeech}                           & Transformer                        & 45.9                        \\
{\color[HTML]{656565} HTS-AT~\cite{chen2022hts}$^{\dagger,\star}$} & {\color[HTML]{656565} Transformer} & {\color[HTML]{656565} 47.1} \\
{\color[HTML]{656565} PaSST~\cite{koutini2021efficient}$^{\dagger,\star}$}  & {\color[HTML]{656565} Transformer} & {\color[HTML]{656565} 47.1} \\ \midrule
KD-CNN                        & CNN                                & 45.4$\pm$0.1                   \\
KD-AST                        & Transformer                        & 47.1$\pm$0.0                   \\ \bottomrule
\end{tabular}
\label{tab:audioset_comp}
\end{table}

\begin{table*}[t]
\centering
\caption{Accuracy of CNN and AST models on ESC-50. $a\rightarrow b$ denotes that the model achieves an accuracy of $a$ and $b$ without and with KD, respectively; $\uparrow$ denotes KD improves the performance.}
\begin{tabular}{@{}cccclclcl@{}}
\toprule
                                                                          &                                                                                             &                                                                                             & \multicolumn{6}{c}{ESC Fine-Tuning KD Teacher Model}                                                                                                                                 \\ \cmidrule(l){4-9} 
\multirow{-2}{*}{\begin{tabular}[c]{@{}c@{}}Student\\ Model\end{tabular}} & \multirow{-2}{*}{\begin{tabular}[c]{@{}c@{}}Student Model\\ AudioSet Pretrain\end{tabular}} & \multirow{-2}{*}{\begin{tabular}[c]{@{}c@{}}KD in Student\\ AudioSet Pretrain\end{tabular}} & \multicolumn{2}{c}{No Pretrain}                 & \multicolumn{2}{c}{AudioSet Pretrain w/o KD}              & \multicolumn{2}{c}{AudioSet Pretrain w/ KD}               \\ \midrule

\multicolumn{3}{r}{{\color[HTML]{656565} \textit{Teacher  (AST-Base) Performance}}}                                                                                                                                                                                   & \multicolumn{2}{c}{{\color[HTML]{656565} \textit{89.23}}} & \multicolumn{2}{c}{{\color[HTML]{656565} \textit{96.07}}} & \multicolumn{2}{c}{{\color[HTML]{656565} \textit{96.45}}} \\ \cmidrule(l){2-9} 
EfficientNet-B2                                                           & No                                                                                          & -                                                                                           & \multicolumn{2}{c}{88.95$\rightarrow$89.58 $\uparrow$}                        & \multicolumn{2}{c}{88.95$\rightarrow$89.80 $\uparrow$}                        & \multicolumn{2}{c}{88.95$\rightarrow$89.87 $\uparrow$}                        \\
EfficientNet-B2                                                           & Yes                                                                                         & No                                                                                          & \multicolumn{2}{c}{96.25$\rightarrow$96.37 $\uparrow$}                        & \multicolumn{2}{c}{96.25$\rightarrow$96.47 $\uparrow$}                        & \multicolumn{2}{c}{96.25$\rightarrow$96.45 $\uparrow$}                        \\
EfficientNet-B2                                                           & Yes                                                                                         & Yes                                                                                         & \multicolumn{2}{c}{96.32$\rightarrow$96.42 $\uparrow$}                        & \multicolumn{2}{c}{96.32$\rightarrow$96.48 $\uparrow$}                        & \multicolumn{2}{c}{96.32$\rightarrow$96.50 $\uparrow$}
                        \\ \midrule
\multicolumn{3}{r}{{\color[HTML]{656565} \textit{Teacher  (EfficientNet-B2) Performance}}}                                                                                                                                                                            & \multicolumn{2}{c}{{\color[HTML]{656565} \textit{88.95}}} & \multicolumn{2}{c}{{\color[HTML]{656565} \textit{96.25}}} & \multicolumn{2}{c}{{\color[HTML]{656565} \textit{96.32}}} \\ \cmidrule(l){2-9} 
AST-Base                                                                  & No                                                                                          & -                                                                                           & \multicolumn{2}{c}{89.23$\rightarrow$89.72 $\uparrow$}                        & \multicolumn{2}{c}{89.23$\rightarrow$90.27 $\uparrow$}                        & \multicolumn{2}{c}{89.23$\rightarrow$90.07 $\uparrow$}                        \\
AST-Base                                                                  & Yes                                                                                         & No                                                                                          & \multicolumn{2}{c}{96.07$\rightarrow$96.08 $\uparrow$}                        & \multicolumn{2}{c}{96.07$\rightarrow$96.22 $\uparrow$}                        & \multicolumn{2}{c}{96.07$\rightarrow$96.02 $\downarrow$}                        \\
AST-Base                                                                  & Yes                                                                                         & Yes                                                                                         & \multicolumn{2}{c}{96.45$\rightarrow$96.57 $\uparrow$}                        & \multicolumn{2}{c}{96.45$\rightarrow$96.67 $\uparrow$}                        & \multicolumn{2}{c}{96.45$\rightarrow$96.80 $\uparrow$}
                        \\ \bottomrule
\end{tabular}
\label{tab:esc50}
\end{table*}

\textbf{CMKD works out-of-the-box.} Though we did not search the KD setting with AudioSet, KD works well on AudioSet for both \texttt{CNN}$\rightarrow$\texttt{AST} and \texttt{AST}$\rightarrow$\texttt{CNN}, demonstrating the proposed cross-model knowledge generalizes for audio classification tasks. In addition, training KD models with more epochs can lead to further performance improvement, which is consistent with our finding that KD models are less prone to overfitting. For example, if we train the AST-Base model with 5 more epochs (10 epochs in total), it achieves an mAP of 47.1, which is 0.3 mAP better than the model trained with 5 epochs. Training no-KD models with more epochs does not lead to such performance improvement.

\textbf{KD leads to larger improvement on smaller datasets}. In all settings, KD is more effective when the model is trained with the smaller balanced training set. This is as expected because AST and CNN models get closer with more training data and cross-model knowledge distillation thus plays a smaller role. However, even with the massive 2M training data, KD still leads to noticeable performance improvement.

\textbf{The advantage of KD narrows after weight averaging and ensemble.} This finding is consistent in FSD50K and AudioSet experiments. We conjecture this is because knowledge distillation limits the \emph{diversity} of model checkpoints of different training epochs and trained with different random seeds as the soft labels from the trained teacher model reduce the student model optimization oscillations.

\textbf{Comparison with State-of-the-art} As a consequence of cross-model knowledge distillation, as shown in Table~\ref{tab:audioset_comp}, our KD-AST model achieves a new state-of-the-art of 47.1$\pm$0.0 mAP on AudioSet, which outperforms the previous state-of-the-art of 45.9 mAP~\cite{gong21b_interspeech}. Our KD-CNN model also outperforms all previous CNN and CNN-attention models. Our KD-AST matches the performance of two concurrent\footnote{\cite{chen2022hts,koutini2021efficient} are arxiv preprints released after we started this work.} efforts~\cite{chen2022hts,koutini2021efficient} that aim to improve the AST architecture and training efficiency. Note that both~\cite{chen2022hts,koutini2021efficient} use a sampling rate of 32kHz, which is higher than the more commonly used and computationally efficient 16kHz sampling rate used in our experiments. In~\cite{kong2020panns}, it is shown that using a 32kHz sampling rate can lead to an improvement of about 0.4 mAP. In other words, our KD-AST matches~\cite{chen2022hts,koutini2021efficient} with lower resolution input. Also, CMKD is orthogonal to model architecture improvement, as we showed in Section~\ref{sec:teacherstudent} and~\ref{sec:hybrid}, CMKD can improve all models we evaluated, thus it can be potentially combined with~\cite{chen2022hts,koutini2021efficient} to further improve model performance.

\subsection{ESC-50 Experiments}
\label{sec:esc50}

Finally, we conduct experiments on ESC-50 to study KD on a tiny dataset and transfer learning. The ESC-50~\cite{piczak2015esc} dataset consists of 2,000 5-second environmental audio recordings organized into 50 classes. We follow the standard 5-fold cross-validation to evaluate our model, repeat each experiment three times, and report the mean accuracy. Note for KD experiments, the teacher model is trained with the same data folds as the student model during cross-validation. 

For both teacher and student models, we consider three settings: 1) model without audio domain pretraining; 2) model pretrained on full AudioSet, but during AudioSet training, KD is not applied; and 3) model pretrained on full AudioSet and CMKD is applied during pretraining. We train or fine-tune the models on ESC-50 with and without KD and report the performance change in Table~\ref{tab:esc50}. In other words, we study the impact of KD in both the pretraining and fine-tuning stages. As with the AudioSet experiments, we re-use the optimal FSD50K KD setting in Table~\ref{tab:kd_setting} but increase the temperature to \{5.0, 6.0\} as the model performance is already high and the prediction of the teacher needs to be further softened. Key findings are as follows:

\textbf{KD leads to small but consistent improvement}: In almost all settings, KD consistently enhances the accuracy of the model. But compared with FSD50K and AudioSet, we find the improvement on ESC-50 is smaller. This is potentially due to the accuracy of the model being already high on ESC-50, thus the distinction between the teacher prediction and ground truth label is small.

\textbf{The advantage of KD in pretraining transfers.} As discussed in Section~\ref{sec:audioset}, models trained with KD have a performance advantage over models trained without KD. When we fine-tuned the AudioSet pretrained model on ESC-50, we find such an advantage still exists. Specifically, no matter if KD is applied in the fine-tuning stage, the model trained with KD in the pretraining stage always performs better. 

Nevertheless, for teacher models, KD in the pretraining stage does not impact its student's performance. This is consistent with our finding in Section~\ref{sec:teacherstudent} that the performance of the teacher is not important. In fact, even for teacher models with and without AudioSet pretraining that have a great performance gap, their student models only have a small performance difference.

The state-of-the-art model~\cite{gong21b_interspeech} achieves 95.6$\pm$0.4\% on ESC-50. Our KD-AST and KD-CNN achieves 96.8$\pm$0.1\% and 96.5$\pm$0.3\%, respectively, both outperform~\cite{gong21b_interspeech}. Two concurrent efforts~\cite{koutini2021efficient,chen2022hts} on improving AST architecture report 96.8\% and 97.0\% on ESC-50. Nevertheless, their results are based on 32kHz audio while our experiment is based on 16kHz audio. Using a higher sampling rate is known to improve the model performance for audio classification~\cite{kong2020panns}, but it could also increase the storage and computation cost.


\section{Related Work}

\textbf{CNN and Transformer} Convolution neural network~\cite{lecun1995convolutional,he2016deep,tan2019efficientnet,howard2017mobilenets} is a class of neural network model that typically has a hierarchical architecture composed of convolution layers, pooling layers, and fully connected layers. In the past decade, convolution neural networks have been the de-facto model for vision and audio tasks. Transformers~\cite{vaswani2017attention} is another class of neural network that is solely based on the self-attention mechanism. It was originally proposed for natural language processing tasks~\cite{devlin-etal-2019-bert,liu2019roberta}, but recently, the Vision Transformer ~\cite{dosovitskiy2020image} and Audio Spectrogram Transformer~\cite{gong21b_interspeech} have been shown to outperform deep learning models constructed with convolutional neural networks (CNNs) on vision and audio tasks, respectively, thus extending the success of Transformers to the vision and audio domain.

Besides pure CNN and pure attention based models, there is another line of research on combining CNN and attention models by integrating convolution operations into attention models, or conversely, introducing attention mechanism into CNN models~\cite{gong_psla,gulati2020conformer,li2021localvit,chen2021mobile,guo2021cmt,xiao2021early,yuan2021incorporating,yan2021contnet,dai2021coatnet,kong2020sound,miyazaki2020convolution}. Unlike these efforts on CNN-attention hybrid models, we show the strength of CNN and Transformer models can be combined more efficiently by simple knowledge distillation without model architecture modification. Both our KD-CNN and KD-AST model outperform the hybrid model for the audio classification tasks. We also show the proposed cross-model knowledge distillation can also be applied to CNN-attention hybrid models.

\textbf{Audio Classification}~\cite{mesaros2017detection} is a task of identifying sound events that occur in a given audio recording. In recent years, audio classification research has moved from small \emph{single-label} datasets such as ESC-50~\cite{piczak2015esc} to much larger \emph{multi-label} datasets such as FSD50K~\cite{fonseca2020fsd50k} and AudioSet~\cite{gemmeke2017audio}. With the support of large and realistic datasets, many new neural network models have been proposed including CNN models (e.g.,~\cite{gemmeke2017audio,kong2020panns,wang2019comparison}), CNN-attention models (e.g.,~\cite{srivastava2021conformer,gong_psla,ford2019deep,miyazaki2020convolution,kong2020sound}), and pure attention based models~\cite{gong21b_interspeech,chen2022hts,koutini2021efficient}. In particular, the pure-attention-based Audio Spectrogram Transformer (AST) and its variants~\cite{gong21b_interspeech,chen2022hts,koutini2021efficient} were shown to outperform CNN and CNN-attention based models and achieved state-of-the-art on multiple audio classification datasets. Unlike these previous efforts, we do not propose a new model architecture but instead, focus on knowledge distillation between these existing models. We show that noticeable performance improvement can be obtained for existing models of all categories (CNN, CNN-attention, and pure attention based models) by simply letting them learn from each other.

\textbf{Knowledge Distillation} was first proposed in~\cite{hinton2015distilling} with the goal of compressing the machine learning model. It refers to the training paradigm in which a \emph{student} model uses predictions from a \emph{teacher} model as soft labels instead of the hard ground truth label as the training target. In the conventional setting, the student model is \emph{smaller} and typically \emph{weaker} than the teacher model. Specifically for the audio classification task, knowledge distillation is mainly used for model compression purpose~\cite{jung2020knowledge,koutini2018iterative,choi2021temporal}. Unlike these previous efforts, our main goal is to improve the model performance with knowledge distillation. We show that a \emph{small} (and weaker) teacher is in fact a good teacher in cross-model knowledge distillation.

Recent studies find knowledge distillation can also be used to transfer the inductive bias of neural networks~\cite{abnar2020transferring,touvron2021training,ren2021co,kuncoro2019scalable}. In particular, it was found that Transformer models can be trained more effectively with CNN teachers for vision tasks~\cite{touvron2021training,ren2021co}. Unlike these previous efforts, in this paper, we focus on the audio domain, and study the \emph{bidirectional} knowledge distillation between CNN and Transformer models rather than just the \texttt{CNN}$\rightarrow$\texttt{Transformer} direction. We find that for both directions, the student model noticeably improves, and in many cases, outperforms the teacher model. With extensive experiments, we discover the optimal knowledge distillation settings for audio classification, which are different from those for the vision tasks.

\section{Conclusion}

In this paper, we show an intriguing interaction between CNN and Transformer models and propose cross-model knowledge distillation for audio classification.  When we use any one of CNN and AST models as the teacher and train another model as the student via knowledge distillation, the student model noticeably improves and outperforms the teacher. With extensive experiments, we find a small model that has a different architecture with the student model is the optimal teacher. For \texttt{CNN}$\rightarrow$\texttt{AST}, using multiple teachers with different architectures can further improve the model performance. Surprisingly, the performance of the teacher model is not important because cross-model knowledge distillation is beyond simple behavior copying. We observe that knowledge distillation makes the model learn faster and be less prone to overfitting as well as improving the model's robustness. Compared with CNN-attention hybrid models, cross-model knowledge distillation is more effective and does not require any model architecture change. As a consequence, our KD-CNN model with only 8M parameters outperforms the original AST with 88M parameters on FSD50K; our KD-AST model performs even better and achieves new state-of-the-art on FSD50K, AudioSet, and ESC50.

\bibliographystyle{IEEEtran}
\bibliography{ref}

\begin{thebibliography}{10}
\providecommand{\url}[1]{#1}
\csname url@samestyle\endcsname
\providecommand{\newblock}{\relax}
\providecommand{\bibinfo}[2]{#2}
\providecommand{\BIBentrySTDinterwordspacing}{\spaceskip=0pt\relax}
\providecommand{\BIBentryALTinterwordstretchfactor}{4}
\providecommand{\BIBentryALTinterwordspacing}{\spaceskip=\fontdimen2\font plus
\BIBentryALTinterwordstretchfactor\fontdimen3\font minus
  \fontdimen4\font\relax}
\providecommand{\BIBforeignlanguage}[2]{{%
\expandafter\ifx\csname l@#1\endcsname\relax
\typeout{** WARNING: IEEEtran.bst: No hyphenation pattern has been}%
\typeout{** loaded for the language `#1'. Using the pattern for}%
\typeout{** the default language instead.}%
\else
\language=\csname l@#1\endcsname
\fi
#2}}
\providecommand{\BIBdecl}{\relax}
\BIBdecl

\bibitem{woodard1992modeling}
J.~P. Woodard, ``Modeling and classification of natural sounds by product code
  hidden markov models,'' \emph{IEEE Transactions on Signal Processing},
  vol.~40, no.~7, pp. 1833--1835, 1992.

\bibitem{goldhor1993recognition}
R.~S. Goldhor, ``Recognition of environmental sounds,'' in \emph{IEEE
  International Conference on Acoustics, Speech, and Signal Processing}, 1993.

\bibitem{chachada2014environmental}
S.~Chachada and C.-C.~J. Kuo, ``Environmental sound recognition: A survey,''
  \emph{APSIPA Transactions on Signal and Information Processing}, vol.~3,
  2014.

\bibitem{lecun1995convolutional}
Y.~LeCun, Y.~Bengio \emph{et~al.}, ``Convolutional networks for images, speech,
  and time series,'' \emph{The handbook of brain theory and neural networks},
  vol. 3361, no.~10, 1995.

\bibitem{hershey2017cnn}
S.~Hershey, S.~Chaudhuri, D.~P. Ellis, J.~F. Gemmeke, A.~Jansen, R.~C. Moore,
  M.~Plakal, D.~Platt, R.~A. Saurous, B.~Seybold \emph{et~al.}, ``Cnn
  architectures for large-scale audio classification,'' in \emph{IEEE
  International Conference on Acoustics, Speech, and Signal Processing}, 2017.

\bibitem{palanisamy2020rethinking}
K.~Palanisamy, D.~Singhania, and A.~Yao, ``Rethinking cnn models for audio
  classification,'' \emph{arXiv preprint arXiv:2007.11154}, 2020.

\bibitem{salamon2017deep}
J.~Salamon and J.~P. Bello, ``Deep convolutional neural networks and data
  augmentation for environmental sound classification,'' \emph{IEEE Signal
  processing letters}, vol.~24, no.~3, pp. 279--283, 2017.

\bibitem{piczak2015environmental}
K.~J. Piczak, ``Environmental sound classification with convolutional neural
  networks,'' in \emph{IEEE International Workshop on Machine Learning for
  Signal Processing}, 2015.

\bibitem{tokozume2017learning}
Y.~Tokozume and T.~Harada, ``Learning environmental sounds with end-to-end
  convolutional neural network,'' in \emph{IEEE International Conference on
  Acoustics, Speech, and Signal Processing}, 2017.

\bibitem{huzaifah2017comparison}
M.~Huzaifah, ``Comparison of time-frequency representations for environmental
  sound classification using convolutional neural networks,'' \emph{arXiv
  preprint arXiv:1706.07156}, 2017.

\bibitem{gong21b_interspeech}
Y.~Gong, Y.-A. Chung, and J.~Glass, ``Ast: Audio spectrogram transformer,'' in
  \emph{Interspeech}, 2021.

\bibitem{koutini2021efficient}
K.~Koutini, J.~Schl{\"u}ter, H.~Eghbal-zadeh, and G.~Widmer, ``Efficient
  training of audio transformers with patchout,'' \emph{arXiv preprint
  arXiv:2110.05069}, 2021.

\bibitem{chen2022hts}
K.~Chen, X.~Du, B.~Zhu, Z.~Ma, T.~Berg-Kirkpatrick, and S.~Dubnov, ``Hts-at: A
  hierarchical token-semantic audio transformer for sound classification and
  detection,'' \emph{arXiv preprint arXiv:2202.00874}, 2022.

\bibitem{vaswani2017attention}
A.~Vaswani, N.~Shazeer, N.~Parmar, J.~Uszkoreit, L.~Jones, A.~N. Gomez,
  L.~Kaiser, and I.~Polosukhin, ``Attention is all you need,'' in
  \emph{Advances in Neural Information Processing Systems}, 2017.

\bibitem{devlin-etal-2019-bert}
J.~Devlin, M.-W. Chang, K.~Lee, and K.~Toutanova, ``{BERT}: Pre-training of
  deep bidirectional transformers for language understanding,'' in
  \emph{Conference of the North American Chapter of the Association for
  Computational Linguistics}, 2019.

\bibitem{dosovitskiy2020image}
A.~Dosovitskiy, L.~Beyer, A.~Kolesnikov, D.~Weissenborn, X.~Zhai,
  T.~Unterthiner, M.~Dehghani, M.~Minderer, G.~Heigold, S.~Gelly, J.~Uszkoreit,
  and N.~Houlsby, ``An image is worth 16x16 words: Transformers for image
  recognition at scale,'' in \emph{International Conference on Learning
  Representations}, 2021.

\bibitem{touvron2021training}
H.~Touvron, M.~Cord, M.~Douze, F.~Massa, A.~Sablayrolles, and H.~J{\'e}gou,
  ``Training data-efficient image transformers \& distillation through
  attention,'' in \emph{International Conference on Machine Learning}, 2021.

\bibitem{raghu2021vision}
M.~Raghu, T.~Unterthiner, S.~Kornblith, C.~Zhang, and A.~Dosovitskiy, ``Do
  vision transformers see like convolutional neural networks?'' \emph{Advances
  in Neural Information Processing Systems}, 2021.

\bibitem{ren2021co}
S.~Ren, Z.~Gao, T.~Hua, Z.~Xue, Y.~Tian, S.~He, and H.~Zhao, ``Co-advise: Cross
  inductive bias distillation,'' \emph{arXiv preprint arXiv:2106.12378}, 2021.

\bibitem{gong_psla}
Y.~Gong, Y.-A. Chung, and J.~Glass, ``Psla: Improving audio tagging with
  pretraining, sampling, labeling, and aggregation,'' \emph{IEEE/ACM
  Transactions on Audio, Speech, and Language Processing}, 2021.

\bibitem{tan2019efficientnet}
M.~Tan and Q.~Le, ``Efficientnet: Rethinking model scaling for convolutional
  neural networks,'' in \emph{International Conference on Machine Learning},
  2019.

\bibitem{8099726}
G.~Huang, Z.~Liu, L.~Van Der~Maaten, and K.~Q. Weinberger, ``Densely connected
  convolutional networks,'' in \emph{IEEE/CVF Conference on Computer Vision and
  Pattern Recognition}, 2017.

\bibitem{gwardys2014deep}
G.~Gwardys and D.~M. Grzywczak, ``Deep image features in music information
  retrieval,'' \emph{International Journal of Electronics and
  Telecommunications}, vol.~60, no.~4, pp. 321--326, 2014.

\bibitem{guzhov2020esresnet}
A.~Guzhov, F.~Raue, J.~Hees, and A.~Dengel, ``Esresnet: Environmental sound
  classification based on visual domain models,'' in \emph{International
  Conference on Pattern Recognition}, 2021.

\bibitem{adapa2019urban}
S.~Adapa, ``Urban sound tagging using convolutional neural networks,'' in
  \emph{DCASE Workshop}, 2019.

\bibitem{liu2022convnet}
Z.~Liu, H.~Mao, C.-Y. Wu, C.~Feichtenhofer, T.~Darrell, and S.~Xie, ``A convnet
  for the 2020s,'' \emph{arXiv preprint arXiv:2201.03545}, 2022.

\bibitem{trockman2022patches}
A.~Trockman and J.~Z. Kolter, ``Patches are all you need?'' \emph{arXiv
  preprint arXiv:2201.09792}, 2022.

\bibitem{hinton2015distilling}
G.~Hinton, O.~Vinyals, and J.~Dean, ``Distilling the knowledge in a neural
  network,'' in \emph{NIPS Deep Learning and Representation Learning Workshop},
  2015.

\bibitem{beyer2021knowledge}
L.~Beyer, X.~Zhai, A.~Royer, L.~Markeeva, R.~Anil, and A.~Kolesnikov,
  ``Knowledge distillation: A good teacher is patient and consistent,''
  \emph{arXiv preprint arXiv:2106.05237}, 2021.

\bibitem{park2019specaugment}
D.~S. Park, W.~Chan, Y.~Zhang, C.-C. Chiu, B.~Zoph, E.~D. Cubuk, and Q.~V. Le,
  ``{SpecAugment}: A simple data augmentation method for automatic speech
  recognition,'' \emph{Interspeech}, 2019.

\bibitem{zhang2018mixup}
H.~Zhang, M.~Cisse, Y.~N. Dauphin, and D.~Lopez-Paz, ``mixup: Beyond empirical
  risk minimization,'' in \emph{International Conference on Learning
  Representations}, 2018.

\bibitem{fonseca2020fsd50k}
E.~Fonseca, X.~Favory, J.~Pons, F.~Font, and X.~Serra, ``Fsd50k: an open
  dataset of human-labeled sound events,'' \emph{arXiv preprint
  arXiv:2010.00475}, 2020.

\bibitem{gemmeke2017audio}
J.~F. Gemmeke, D.~P. Ellis, D.~Freedman, A.~Jansen, W.~Lawrence, R.~C. Moore,
  M.~Plakal, and M.~Ritter, ``Audio set: An ontology and human-labeled dataset
  for audio events,'' in \emph{IEEE International Conference on Acoustics,
  Speech, and Signal Processing}, 2017.

\bibitem{piczak2015esc}
K.~J. Piczak, ``Esc: Dataset for environmental sound classification,'' in
  \emph{ACM International Conference on Multimedia}, 2015.

\bibitem{shah2018closer}
A.~Shah, A.~Kumar, A.~G. Hauptmann, and B.~Raj, ``A closer look at weak label
  learning for audio events,'' \emph{arXiv preprint arXiv:1804.09288}, 2018.

\bibitem{meire2019impact}
M.~Meire, L.~Vuegen, and P.~Karsmakers, ``The impact of missing labels and
  overlapping sound events on multi-label multi-instance learning for sound
  event classification,'' in \emph{DCASE Workshop}, 2019.

\bibitem{fonseca2020addressing}
E.~Fonseca, S.~Hershey, M.~Plakal, D.~P. Ellis, A.~Jansen, and R.~C. Moore,
  ``Addressing missing labels in large-scale sound event recognition using a
  teacher-student framework with loss masking,'' \emph{IEEE Signal Processing
  Letters}, vol.~27, pp. 1235--1239, 2020.

\bibitem{xie2020self}
Q.~Xie, M.-T. Luong, E.~Hovy, and Q.~V. Le, ``Self-training with noisy student
  improves imagenet classification,'' in \emph{IEEE/CVF Conference on Computer
  Vision and Pattern Recognition}, 2020.

\bibitem{raghu2017svcca}
M.~Raghu, J.~Gilmer, J.~Yosinski, and J.~Sohl-Dickstein, ``Svcca: Singular
  vector canonical correlation analysis for deep learning dynamics and
  interpretability,'' \emph{Advances in Neural Information Processing Systems},
  2017.

\bibitem{naseer2021intriguing}
M.~M. Naseer, K.~Ranasinghe, S.~H. Khan, M.~Hayat, F.~Shahbaz~Khan, and M.-H.
  Yang, ``Intriguing properties of vision transformers,'' \emph{Advances in
  Neural Information Processing Systems}, 2021.

\bibitem{abnar2020transferring}
S.~Abnar, M.~Dehghani, and W.~Zuidema, ``Transferring inductive biases through
  knowledge distillation,'' \emph{arXiv preprint arXiv:2006.00555}, 2020.

\bibitem{srivastava2021conformer}
S.~Srivastava, Y.~Wang, A.~Tjandra, A.~Kumar, C.~Liu, K.~Singh, and Y.~Saraf,
  ``Conformer-based self-supervised learning for non-speech audio tasks,''
  \emph{arXiv preprint arXiv:2110.07313}, 2021.

\bibitem{ford2019deep}
L.~Ford, H.~Tang, F.~Grondin, and J.~R. Glass, ``A deep residual network for
  large-scale acoustic scene analysis.'' in \emph{Interspeech}, 2019.

\bibitem{miyazaki2020convolution}
K.~Miyazaki, T.~Komatsu, T.~Hayashi, S.~Watanabe, T.~Toda, and K.~Takeda,
  ``Convolution augmented transformer for semi-supervised sound event
  detection,'' in \emph{DCASE Workshop}, 2020.

\bibitem{kong2020sound}
Q.~Kong, Y.~Xu, W.~Wang, and M.~D. Plumbley, ``Sound event detection of weakly
  labelled data with cnn-transformer and automatic threshold optimization,''
  \emph{IEEE/ACM Transactions on Audio, Speech, and Language Processing},
  vol.~28, pp. 2450--2460, 2020.

\bibitem{wu2021wav2clip}
H.-H. Wu, P.~Seetharaman, K.~Kumar, and J.~P. Bello, ``Wav2clip: Learning
  robust audio representations from clip,'' \emph{arXiv preprint
  arXiv:2110.11499}, 2021.

\bibitem{YAMNet}
``Sound classification with yamnet,''
  \url{https://www.tensorflow.org/hub/tutorials/yamnet}.

\bibitem{verma2021audio}
P.~Verma and J.~Berger, ``Audio transformers: Transformer architectures for
  large scale audio understanding. adieu convolutions,'' \emph{arXiv preprint
  arXiv:2105.00335}, 2021.

\bibitem{fonseca2021improving}
E.~Fonseca, A.~Ferraro, and X.~Serra, ``Improving sound event classification by
  increasing shift invariance in convolutional neural networks,'' \emph{arXiv
  preprint arXiv:2107.00623}, 2021.

\bibitem{izmailov2018averaging}
P.~Izmailov, D.~Podoprikhin, T.~Garipov, D.~Vetrov, and A.~G. Wilson,
  ``Averaging weights leads to wider optima and better generalization,'' in
  \emph{Conference on Uncertainty in Artificial Intelligence}, 2018.

\bibitem{kong2020panns}
Q.~Kong, Y.~Cao, T.~Iqbal, Y.~Wang, W.~Wang, and M.~D. Plumbley, ``Panns:
  Large-scale pretrained audio neural networks for audio pattern recognition,''
  \emph{IEEE/ACM Transactions on Audio, Speech, and Language Processing},
  vol.~28, pp. 2880--2894, 2020.

\bibitem{wang2019comparison}
Y.~Wang, J.~Li, and F.~Metze, ``A comparison of five multiple instance learning
  pooling functions for sound event detection with weak labeling,'' in
  \emph{IEEE International Conference on Acoustics, Speech, and Signal
  Processing}, 2019.

\bibitem{koutini2018iterative}
K.~Koutini, H.~Eghbal-zadeh, and G.~Widmer, ``Iterative knowledge distillation
  in r-cnns for weakly-labeled semi-supervised sound event detection,'' in
  \emph{DCASE Workshop}, 2018.

\bibitem{he2016deep}
K.~He, X.~Zhang, S.~Ren, and J.~Sun, ``Deep residual learning for image
  recognition,'' in \emph{IEEE/CVF Conference on Computer Vision and Pattern
  Recognition}, 2016.

\bibitem{howard2017mobilenets}
A.~G. Howard, M.~Zhu, B.~Chen, D.~Kalenichenko, W.~Wang, T.~Weyand,
  M.~Andreetto, and H.~Adam, ``Mobilenets: Efficient convolutional neural
  networks for mobile vision applications,'' \emph{arXiv preprint
  arXiv:1704.04861}, 2017.

\bibitem{liu2019roberta}
Y.~Liu, M.~Ott, N.~Goyal, J.~Du, M.~Joshi, D.~Chen, O.~Levy, M.~Lewis,
  L.~Zettlemoyer, and V.~Stoyanov, ``Roberta: A robustly optimized bert
  pretraining approach,'' \emph{arXiv preprint arXiv:1907.11692}, 2019.

\bibitem{gulati2020conformer}
A.~Gulati, J.~Qin, C.-C. Chiu, N.~Parmar, Y.~Zhang, J.~Yu, W.~Han, S.~Wang,
  Z.~Zhang, Y.~Wu \emph{et~al.}, ``Conformer: Convolution-augmented transformer
  for speech recognition,'' \emph{arXiv preprint arXiv:2005.08100}, 2020.

\bibitem{li2021localvit}
Y.~Li, K.~Zhang, J.~Cao, R.~Timofte, and L.~Van~Gool, ``Localvit: Bringing
  locality to vision transformers,'' \emph{arXiv preprint arXiv:2104.05707},
  2021.

\bibitem{chen2021mobile}
Y.~Chen, X.~Dai, D.~Chen, M.~Liu, X.~Dong, L.~Yuan, and Z.~Liu,
  ``Mobile-former: Bridging mobilenet and transformer,'' \emph{arXiv preprint
  arXiv:2108.05895}, 2021.

\bibitem{guo2021cmt}
J.~Guo, K.~Han, H.~Wu, C.~Xu, Y.~Tang, C.~Xu, and Y.~Wang, ``Cmt: Convolutional
  neural networks meet vision transformers,'' \emph{arXiv preprint
  arXiv:2107.06263}, 2021.

\bibitem{xiao2021early}
T.~Xiao, P.~Dollar, M.~Singh, E.~Mintun, T.~Darrell, and R.~Girshick, ``Early
  convolutions help transformers see better,'' \emph{Advances in Neural
  Information Processing Systems}, 2021.

\bibitem{yuan2021incorporating}
K.~Yuan, S.~Guo, Z.~Liu, A.~Zhou, F.~Yu, and W.~Wu, ``Incorporating convolution
  designs into visual transformers,'' in \emph{IEEE/CVF International
  Conference on Computer Vision}, 2021.

\bibitem{yan2021contnet}
H.~Yan, Z.~Li, W.~Li, C.~Wang, M.~Wu, and C.~Zhang, ``Contnet: Why not use
  convolution and transformer at the same time?'' \emph{arXiv preprint
  arXiv:2104.13497}, 2021.

\bibitem{dai2021coatnet}
Z.~Dai, H.~Liu, Q.~Le, and M.~Tan, ``Coatnet: Marrying convolution and
  attention for all data sizes,'' \emph{Advances in Neural Information
  Processing Systems}, 2021.

\bibitem{mesaros2017detection}
A.~Mesaros, T.~Heittola, E.~Benetos, P.~Foster, M.~Lagrange, T.~Virtanen, and
  M.~D. Plumbley, ``Detection and classification of acoustic scenes and events:
  Outcome of the dcase 2016 challenge,'' \emph{IEEE/ACM Transactions on Audio,
  Speech, and Language Processing}, vol.~26, no.~2, pp. 379--393, 2017.

\bibitem{jung2020knowledge}
J.-W. Jung, H.-S. Heo, H.-J. Shim, and H.-J. Yu, ``Knowledge distillation in
  acoustic scene classification,'' \emph{IEEE Access}, vol.~8, pp.
  166\,870--166\,879, 2020.

\bibitem{choi2021temporal}
K.~Choi, M.~Kersner, J.~Morton, and B.~Chang, ``Temporal knowledge distillation
  for on-device audio classification,'' \emph{arXiv preprint arXiv:2110.14131},
  2021.

\bibitem{kuncoro2019scalable}
A.~Kuncoro, C.~Dyer, L.~Rimell, S.~Clark, and P.~Blunsom, ``Scalable
  syntax-aware language models using knowledge distillation,'' \emph{arXiv
  preprint arXiv:1906.06438}, 2019.

\end{thebibliography}

\begin{IEEEbiography}
    [{\includegraphics[width=1in,height=1.25in,clip,keepaspectratio]{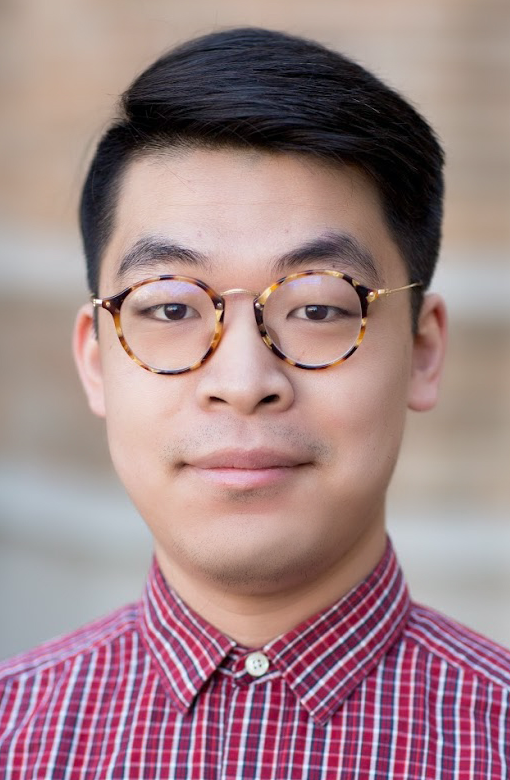}}]{Yuan Gong}
 (Member, IEEE) received the B.S. degree in biomedical engineering from Fudan University, Shanghai, China, and the Ph.D. degree in computer science from the University of Notre Dame, IN, USA, in 2015 and 2020, respectively. He is a Postdoctoral Researcher with the MIT Computer Science and Artificial Intelligence Laboratory. Currently, his primary research interests include automatic speech recognition, speech based healthcare system, and acoustic events detection. He won the 2017 AVEC depression detection challenge and one of his papers was nominated for the Best Student Paper Award in Interspeech 2019.
\end{IEEEbiography}

\begin{IEEEbiography}
    [{\includegraphics[width=1in,height=1.25in,clip,keepaspectratio]{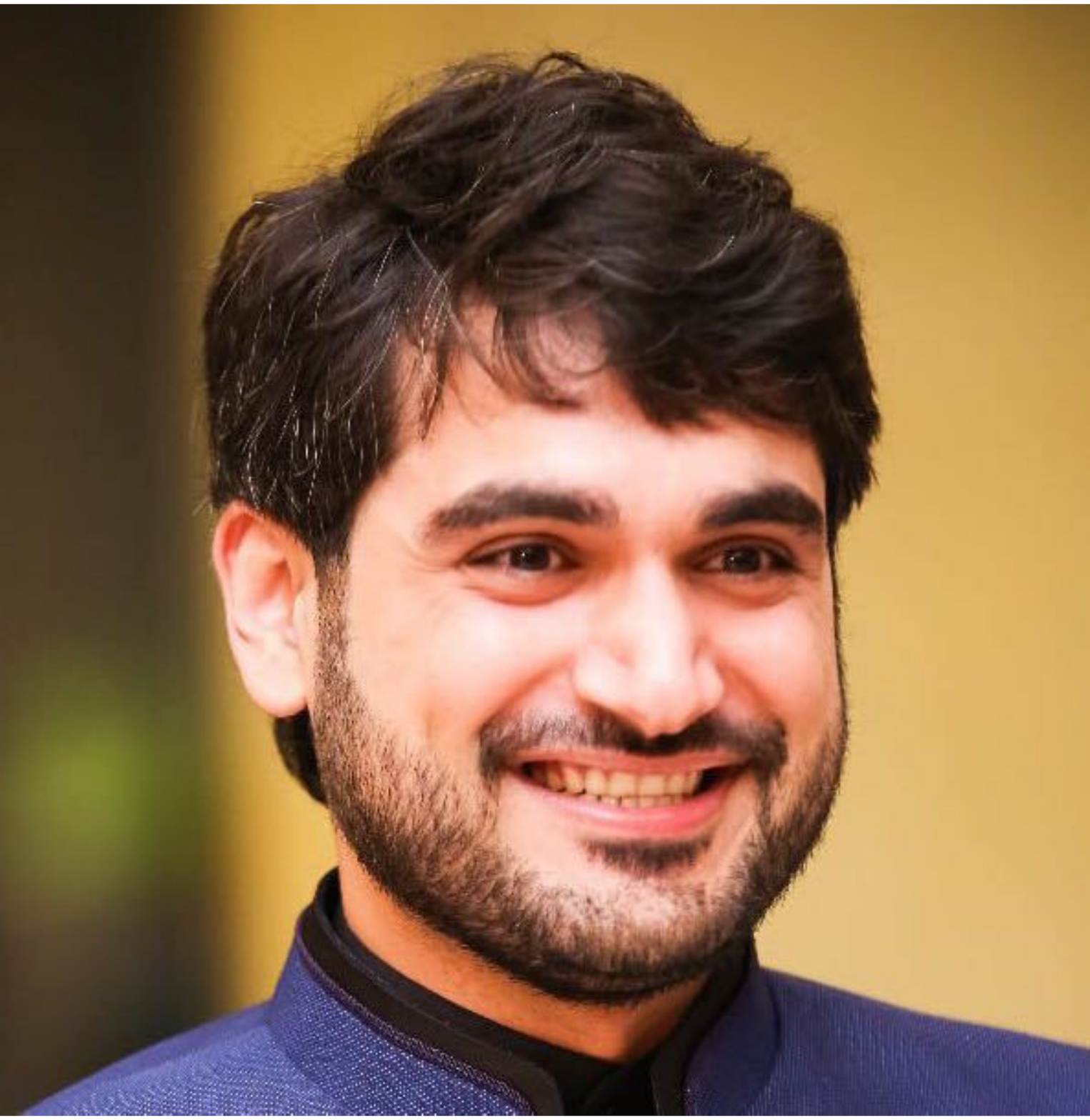}}]{Sameer Khurana}
is a Ph.D. candidate in Electrical Engineering and Computer Science at MIT. He works with Dr. James Glass in MIT's Computer Science and Artificial Intelligence Lab on speech recognition, translation, and self-supervised speech processing. His primary research interest is to build automatic speech recognition and translation models that could learn from a few human-labeled examples. In addition, his work aims to expand current automatic speech recognition and translation technology to low-resource languages.
\end{IEEEbiography}

\begin{IEEEbiography}
    [{\includegraphics[width=1in,height=1.25in,clip,keepaspectratio]{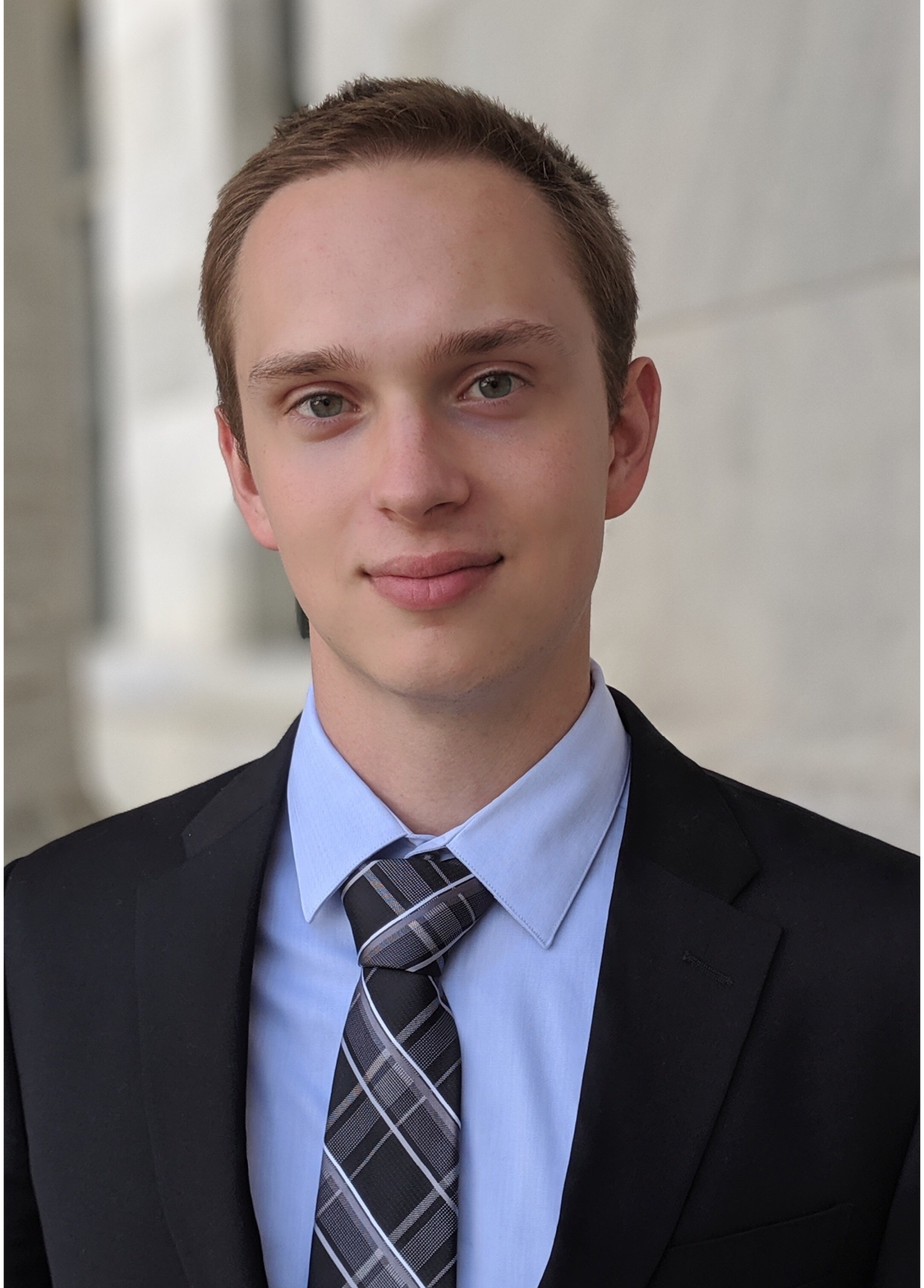}}]{Andrew Rouditchenko}
received the S.B. degree in electrical engineering and computer science from MIT in 2019 and the M.Eng degree in electrical engineering and computer science from MIT in 2021. He is a Ph.D. student working with Dr. Jim Glass in the MIT Spoken Language Systems Group.
\end{IEEEbiography}

\begin{IEEEbiography}
    [{\includegraphics[width=1in,height=1.25in,clip,keepaspectratio]{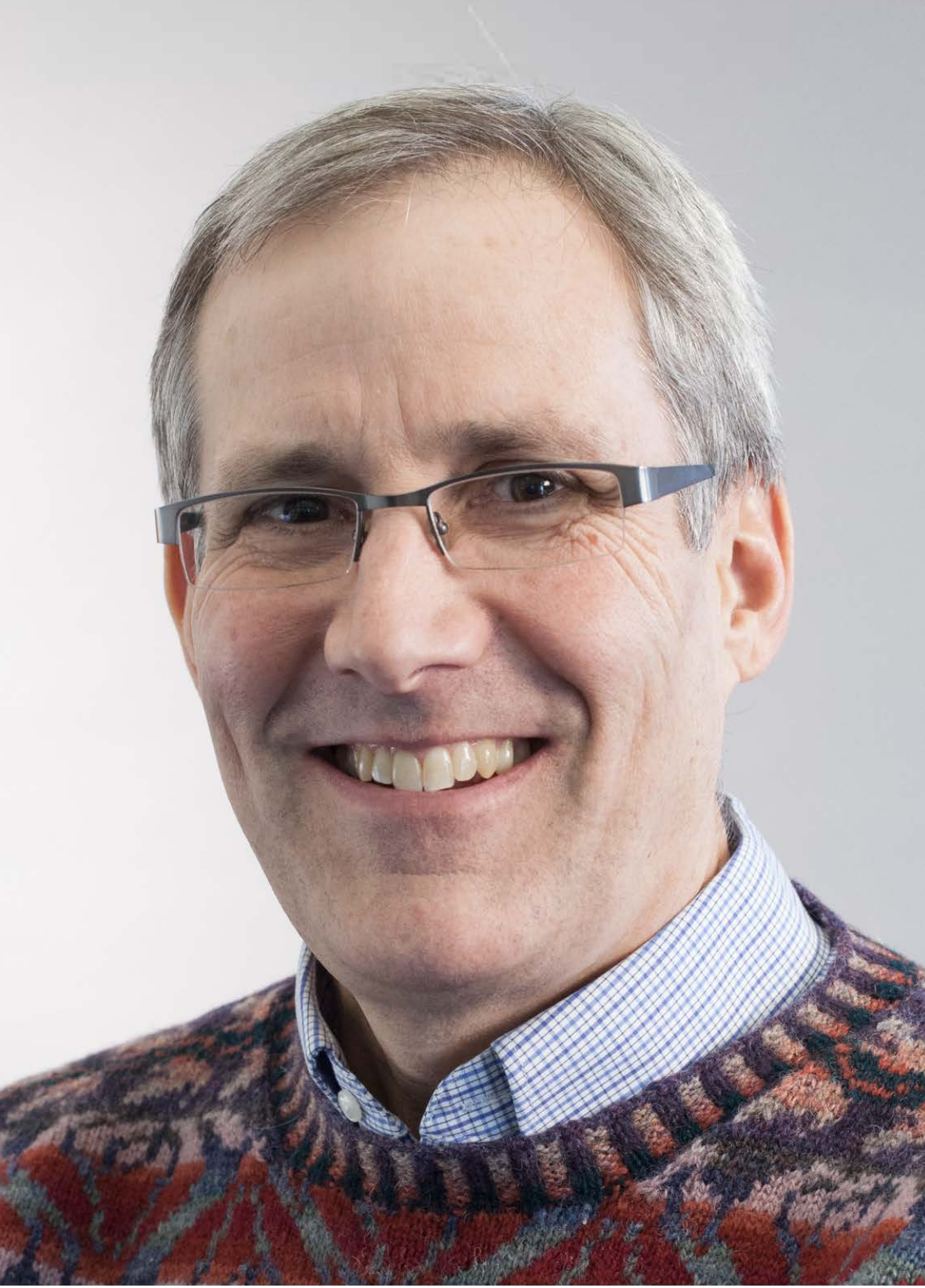}}]{James Glass}
(Fellow, IEEE) is a Senior Research Scientist with MIT where he Leads the Spoken Language Systems Group with the Computer Science and Artificial Intelligence Laboratory. He is also a member of the Harvard University Program in Speech and Hearing Bioscience and Technology. Since obtaining the S.M. and Ph.D. degrees with MIT in electrical engineering and computer science, his research interests include automatic speech recognition, unsupervised speech processing, and spoken language understanding. He is a Fellow of the International Speech Communication Association, and is currently an Associate Editor for the IEEE Transactions on Pattern Analysis and Machine Intelligence.
\end{IEEEbiography}

\end{document}